\documentclass[aps,prd,twocolumn,preprintnumbers,floatfix,nofootinbib,showpacs]{revtex4-1}
\usepackage{graphicx}
\usepackage{bm}
\usepackage{times}
\usepackage{amsmath}
\usepackage{slashed}
\usepackage{color}
\usepackage{natbib}
\usepackage{color}
\usepackage{hyperref}
\usepackage[]{appendix}
\usepackage{multirow}
\usepackage{tabularx}

\newcommand{\alf}{Alfv\'en }
\newcommand{\rev}[1]{\textcolor{black} {#1}}
\begin{document}

\title{Cosmic Ray Protons in the Inner Galaxy and the Galactic Center Gamma-Ray Excess}

\author{Eric Carlson}\email{erccarls@ucsc.edu}   
\author{Stefano Profumo}\email{profumo@ucsc.edu}

\affiliation{Department of Physics and Santa Cruz Institute for Particle Physics
University of California, Santa Cruz, CA 95064, USA
}

\pacs{95.85.Pw, 96.50.S-, 96.50.sb, 95.35.+d}
\date{\today}
\vspace{1cm}

\begin{abstract}
A gamma-ray excess over background has been claimed in the inner regions of the Galaxy, triggering some excitement about the possibility that the gamma rays originate from the annihilation of dark matter particles. We point out that the existence of such an excess depends on how the diffuse gamma-ray background is defined, and on the procedure employed to fit such background to observations. We demonstrate that a gamma-ray emission with spectral and morphological features closely matching the observed excess arises from a population of cosmic ray protons in the inner Galaxy, and provide proof of principle and arguments for the existence of such a population, most likely originating from local supernova remnants. Specifically, the ``Galactic center excess'' is readily explained by a recent cosmic-ray injection burst, with an age in the 1-10 kilo-year range, while the extended inner Galaxy excess points to mega-year old injection episodes, continuous or impulsive. We conclude that it is premature to argue that there are no standard astrophysical mechanisms that can explain the excess. 
\end{abstract}

\maketitle

\section{Introduction}

The Galactic center is a promising location to search for non-gravitational signals from particle dark matter such as gamma rays from dark matter pair annihilation. Any model for the density distribution of dark matter in the Galaxy predicts a high concentration of dark matter in the Galactic center, with a resulting large number density of dark matter particle pairs. Barring the possibility of a large, nearby dark matter ``clump'', the Galactic center direction is the direction in the sky where the line-of-sight integral of the dark matter density squared is maximal. As a result, the Galactic center is the location where one of the brightest photon signals from dark matter annihilation is expected.

On the downside, the center of the Galaxy hosts a complex combination of ``standard'' astrophysical $\gamma$-ray sources. The region contains numerous resolved and many unresolved $\gamma$-ray point sources; in addition, the diffuse Galactic emission is brightest in the center of the Galaxy, where the largest macroscopic concentrations of gas, cosmic rays and interstellar radiation energy density are found. This dense environment copiously sources $\gamma$ rays from hadronic inelastic interactions as well as from inverse Compton scattering and bremsstrahlung. Such complex background structure can be hardly reconstructed from first principles, and non-trivial extrapolations and inference often, if not always, define how the predicted background emission is calculated.

The combination of such an appealing target with such a treacherous background has contributed to much debate about the existence and nature of excess $\gamma$-ray emission from the Galactic center region. Ever since the years of the EGRET telescope, claims of an excess diffuse $\gamma$-ray emission (extending even beyond the Galactic center) have been made \cite{Hunger:1997we, deBoer:2005tm}, and proved premature, with several groups proposing a Galactic cosmic-ray spectra differing from local values~\cite{Strong:2004de, Kamae:2004xx}.  The EGRET excess was subsequently shown to be systematic in origin, relating instead to a miscalculation of EGRET's sensitivity above a few GeV~\cite{Stecker:2008}, and was later shown to be conclusively unfounded \cite{Abdo:2010nz} using data from the Fermi Large Area Telescope (LAT) \cite{Atwood:2009ez}.

Shortly after LAT data were made public, claims of a Galactic center Excess (GCE) have been put forward, pointing to differing particle dark matter properties (including the preferred mass, pair-annihilation rate, and  annihilation pathway) depending on the background model employed in the analysis, see e.g. \cite{Goodenough:2009gk, Hooper:2010mq, Hooper:2011ti}. 

Several immediate issues have been raised following the identification of excess of $\gamma$ rays over background and with associations to new physics.  These include the question of $\gamma$-ray point source modeling associated with the radio source Sgr A*, see e.g. \cite{Boyarsky:2010dr, Chernyakova:2011zz, Linden:2012bp, Linden:2012iv}, and the role of unresolved populations of $\gamma$-ray emitters such as millisecond pulsars \cite{2011JCAP...03..010A} (see however \cite{Hooper:2013nhl}).

One of the key elements in assessing the presence of a genuine $\gamma$-ray excess in the Galactic center region is, naturally, that of modeling $\gamma$-ray sources in the region. Critical to this is the role of unidentified point sources, including population models for unidentified source classes, and of sources whose spectrum and even source extension is unclear (for example the $\gamma$-ray counterpart to Sgr A*). A second key element is the diffuse $\gamma$-ray emission induced by Galactic cosmic rays. It has long been known \cite{Stecker1977a, Abdo:2010nz} that the key components of such emission, in the 0.1-100 GeV range are (i) hadronic emission from neutral pion decay produced by inelastic proton collision with the interstellar gas, (ii) inverse Compton up-scattering of background interstellar radiation by cosmic-ray electrons and positrons, and (iii) bremsstrahlung. 

We review below how the two key ingredients to the background model employed to extract the Galactic center excess have been handled in the three most recent and comprehensive analyses. What we believe is a crucial point to make is that the general procedure, in those studies, has been to employ background templates that make crucial assumptions about the Galactic diffuse emission. One of us had pointed out in Ref.~\cite{Linden:2010ea}, with Linden, that important systematic effects in extracting a diffuse $\gamma$-ray excess originate from neglecting the cosmic-ray density distribution and in utilizing templates where the diffuse hadronic emission, item (i) in the list above, follows the morphology of the target gas density. In the present study, we point out that (a) very little is known about cosmic rays in the Galactic center region; that (b) more or less young populations of cosmic rays are likely to inhabit that region and to importantly contribute to the hadronic emission in a way that would be completely missed by a current template analysis; and, finally, that (c) such a population(s) is likely to source the claimed $\gamma$-ray excess.

Let us first briefly review three recent studies devoted to the Galactic center excess, Ref.~\cite{gordon_macias:2013}, \cite{Abazajian:2014fta} and \cite{Daylan:2014rsa}. The study presented in Ref.~\cite{gordon_macias:2013} focuses on the $7^\circ\times7^\circ$ region centered around the Galactic center (GC, $b=0,\ l=0$), and employs the recommended LAT Collaboration diffuse background model {\tt gal\_2yearp7v6\_v0} (we will comment below on the implicit assumptions included in this model), plus isotropic backgrounds, and known $\gamma$-ray sources in the second year Fermi catalogue (2FGL). The study confirms evidence for a spherically symmetric extended source, as obtained in previous studies \cite{Abazajian:2012pn}, with a spectrum consistent both with emission from millisecond pulsars and with dark matter annihilation. Ref.~\cite{gordon_macias:2013} also attempts to assess systematic uncertainties in the background modeling, concluding that such uncertainty is in the vicinity of the 20\% level.  \rev{In a follw-up paper~\cite{gordan2014} by the same authors, 20 cm templates tracing the molecular gas distribution were added to the likelihood analysis and were found to significantly improve the fit while still robustly detecting an approximately spherically symmetric GCE counterpart.}

The analysis of Ref.~\cite{Abazajian:2014fta} also considers a region of interest of $7^\circ\times7^\circ$  centered around the GC, and employs two choices for the energy range, photon source class, pixel size, and energy binning. Ref.~\cite{Abazajian:2014fta} then fits a variety of templates to the observed $\gamma$-ray data. These templates include, in addition to point sources, the recommended Galactic diffuse emission model {\tt gal\_2yearp7v6\_v0} and isotropic background model {\tt iso\_p7v6source}, a template (MG) that intends to map the bremsstrahlung emission associated with high-energy electrons interacting with molecular gas clouds as traced by the 20 cm radio map of the GC \cite{YusefZadeh:2012nh}, a Galactic Center Excess (GCE) source, and a ``new diffuse'' component associated with a central stellar cluster, with varying spatial profiles.

The two key findings of Ref.~\cite{Abazajian:2014fta} are that (i) an extended emission in the GC region associated with the GCE template is present with any combination of templates and with both choices for the pixel and energy binning etc.; and that (ii) the fluxes and spectra associated with both the $\gamma$-ray emission from the central point source Sgr A* and with the GC extended emission are significantly affected by the choice of the background model, especially in the low-energy range. The GC excess is found to have a spatial distribution consistent with a profile $\propto r^{-2.2}$.

The study of Ref.~\cite{Daylan:2014rsa}, which appeared less than 10 days after Ref.~\cite{Abazajian:2014fta}, focused on an ``Inner Galaxy'' region, which masks out the Galactic plane ($|b|<1^\circ$) and includes a large region of several tens of degrees, and on a ``Galactic center'' region, defined by $|b|<5^\circ$ and $|l|<5^\circ$. Both studies use a novel cut on photon events based on the CTBCORE variable, producing higher resolution maps. In the ``Inner Galaxy'' analysis, Ref.~\cite{Daylan:2014rsa} makes use of three templates (the Fermi collaboration {\tt p6v11} Galactic diffuse model, an isotropic background and a uniform-brightness template matching the Fermi bubbles) plus a ``dark matter'' template of variable inner slope. In the ``Galactic center'' analysis, the templates used include a Galactic diffuse emission  provided by the Fermi collaboration ({\tt gal\_2yearp7v6\_v0}, the same choice as Ref.~\cite{Abazajian:2014fta}), a template tracing the 20 cm emission, along the lines of Ref.~\cite{Abazajian:2014fta}, an isotropic component, and all 2FGL point sources \cite{Fermi-LAT:2011iqa}. As in Ref.~\cite{Abazajian:2014fta} it is found that the isotropic component needed to provide an optimal fit is considerably brighter than the extragalactic $\gamma$-ray background. 

Ref.~\cite{Daylan:2014rsa} indicates a strong preference for the existence of a Galactic center excess, and finds a similar preferred spatial distribution profile to Ref.~\cite{Abazajian:2014fta} and, generically, a similar preferred spectral shape. Ref.~\cite{Daylan:2014rsa} points out that the excess is approximately spherically symmetric. From both spectral and morphological considerations, Ref.~\cite{Daylan:2014rsa} argues that a population of unresolved millisecond pulsars (MSP) in the relevant Galactic region is strongly disfavored. Also, as pointed out in Ref.~\cite{Linden:2010ea}, based on the population of {\em resolved} MSPs, the contribution from an unresolved population should account for less than $\sim5-10$ \% of the $\gamma$-ray excess (see also \cite{Hooper:2013nhl}).

It is apparent that a central issue to the determination of the existence of any diffuse $\gamma$-ray excess is whether or not the background model for the Galactic diffuse emission accurately reproduces the expected $\gamma$-ray emission. All recent studies reviewed above employ a diffuse Galactic model recommended by the Fermi collaboration for use with Pass 7 LAT data \cite{diffusemodel}. Interestingly, the Collaboration explicitly (and in bold face) discourages the use of one the most recent such model for Pass 7 reprocessed data ``for analyses of spatially extended sources in the region defined in Fig.~1'', a region which includes the Galactic center region (as noted in Ref.~\cite{Daylan:2014rsa}). While the key concern is the inclusion, in the reprocessed data background model, of sources with extension more than 2$^\circ$, it is also apparent that such background models are not designed with the purpose of establishing the existence of a diffuse emission.

One of the key issues with using the diffuse model recommended by the Fermi Collaboration for the purposes of establishing a diffuse excess is the set of templates employed to reproduce the morphology of the hadronic and inverse-Compton Galactic diffuse emission. Employing gas column-density map templates to reproduce the diffuse $\gamma$-ray intensity entirely neglects the possibility of a significantly enhanced cosmic-ray abundance in the inner Galaxy, which almost certainly exists. Similarly, the inverse-Compton template is based, and sensitively depends, on specific choices for the input parameters in the {\tt Galprop} code, most significantly source distribution, diffusive halo geometry and source spectrum (see e.g. \cite{CASANDJIAN}). 

Other quite relevant issues with the Fermi Collaboration recommended diffuse model have been discussed and tackled in the recent studies of Ref.~\cite{Abazajian:2014fta} and \cite{Daylan:2014rsa}. These include a component of bremsstrahlung emission corresponding, and traced by, molecular gas \cite{Abazajian:2014fta, Daylan:2014rsa}; a diffuse component with a density profile tracing the Milky Way Central Stellar Cluster \cite{Abazajian:2014fta}; and the so-called Fermi bubbles \cite{Daylan:2014rsa}, whose intensity however quite likely deviates from the uniform-brightness assumption of Ref.~\cite{Daylan:2014rsa}.

With all the mentioned caveat in mind, in the present study we show that simple Galactic cosmic-ray models exist that naturally explain the observed excess. The origin of such cosmic rays is likely associated either with supernova remnants in the inner Galactic region, or with past activity of Sgr A*, or both. We demonstrate that there is no spectral or morphological preference for dark matter over such cosmic-ray models, whose existence in the inner Galaxy is more than plausible. Based on Occam's razor principle, we argue that the Galactic center excess finds a much more compelling interpretation in the context of cosmic-ray models for the inner Galaxy rather than in that of dark matter annihilation.

\section{Cosmic-ray protons in the inner Galaxy}

\subsection{Morphological properties}
\label{sec:morphology}

There exist two key potential sources of cosmic rays in the inner Galaxy within the energy range relevant here: (i) supernova remnants and (ii) past activity of the central supermassive black hole associated with the radio source Sgr A*. For simplicity, we assume that both sources injected cosmic rays at the center of the Galaxy ($l=0,\ b=0$) at one or more points in time in the past. We will assume both an impulsive and a continuous injection for the sources, the former arguably more plausible for Sgr A* or for isolated star-formation bursts, and the latter closer to what expected for a population of supernova remnants. We first feature a qualitative analytic discussion (sec.~\ref{sec:ana}), and we then present detailed results obtained with a full cosmic-ray propagation simulation with the {\tt Galprop} package (sec.~\ref{sec:nummorph}).

\subsubsection{Analytic Estimates}\label{sec:ana}

In the case of an impulsive source, the spatial distribution of the protons after a time $t_i$ can be approximated  as follows \cite{aharonian}:
\begin{equation}\label{eq:imp}
f(r)\propto\frac{\exp[-r^2/R_{\rm dif}^2(t_i)]}{R_{\rm dif}^3(t_i)},
\end{equation}
where the diffusion radius
\begin{equation}\label{eq:rdif}
R_{\rm dif}(E,t)=2\sqrt{D(E)t\frac{\exp[t\delta/\tau_{\rm pp}]-1}{t\delta/\tau_{\rm pp}}},
\end{equation}
with $D(E)=D_0(E/4~\rm{GeV})^\delta$, and where $\tau_{\rm pp}$ is the approximately energy-independent proton cooling time. For timescales $t\ll\tau_{\rm pp}$, $R_{\rm dif}\approx2\sqrt{D(E)t}$, indicative of a purely Brownian process. Note that the particle spectrum clearly depends on position unless $\delta=0$, since the quantity $R_{\rm dif}$, driving the spatial dependence, depends on energy. In the narrowly peaked energy range of interest for the Galactic Center excess, such effect is, however, limited. For example, to first order and for $\delta\sim0.33$, $R_{\rm dif}\propto E^{0.16}$ which is less than a factor 1.5 difference from 10 GeV to 100 GeV.

The counterpart to Eq.~(\ref{eq:imp}) for a continuous source is given by 
\begin{equation}
f(r)\propto\frac{{\rm erfc}[r/R_{\rm dif}]}{r},
\end{equation}
where erfc is the error-function, and with the same $R_{\rm dif}$ as in Eq.~(\ref{eq:rdif}) but with $t$, this time, referring to the time at which the continuous cosmic-ray source started injecting particles. In the limit $r\ll R_{\rm dif}$, the cosmic-ray flux saturates to a density $\propto1/r$.

For a diffusion coefficient $D(E)=D_0(E_{\rm p}/4\ {\rm GeV})^\delta$, with $D_0=6.1 \times 10^{28}\ {\rm cm}^2{\rm s}^{-1}$, we then consider a variety of impulsive and continuous sources, with ages listed in Table~\ref{tab:sources} along with their physical and angular diffusion radii at 2~GeV, where the GCE peaks.
\newcolumntype{C}[1]{>{\centering\let\newline\\\arraybackslash\hspace{0pt}}m{#1}}

\begin{table}[t]
  \squeezetable
  \centering
  \begin{tabular}{C{1.5cm} C{1.5cm} C{1.5cm} C{1.5cm} C{1.5cm}}
	\hline \hline
	\multicolumn{1}{c}{Name} & \multicolumn{1}{c}{Type} & \multicolumn{1}{c}{Age} &  \multicolumn{2}{c}{$R_{\rm diff}(2~{\rm GeV})$} \\ \hline
    Im1  & Impulse & .5 Kyr & 18 pc & $0.12^\circ$  \\ 
    Im2  & Impulse & 2.5 Kyr & 40 pc & $0.28^\circ$ \\ 
    Im3  & Impulse & 19 Kyr & 110 pc & $0.76^\circ$ \\ 
    Im4  & Impulse & 100 Kyr & 250 pc & $1.7^\circ$ \\ 
    Im5  & Impulse & 2 Myr & 1.13 Kpc & $7.8^\circ$ \\ 
    C1  & Continuous & 7.5 Myr & 2.19 pc & $15^\circ$ \\ 
    C2  & Continuous & $\gtrsim 1$~Gyr & $\infty$ & $\infty$ \\
    \hline
  \end{tabular}
    \caption{Properties of a few benchmark emission sources.}
     \label{tab:sources}
\end{table}

\begin{figure}[t]
\begin{centering}
		\includegraphics[width=0.9\columnwidth]{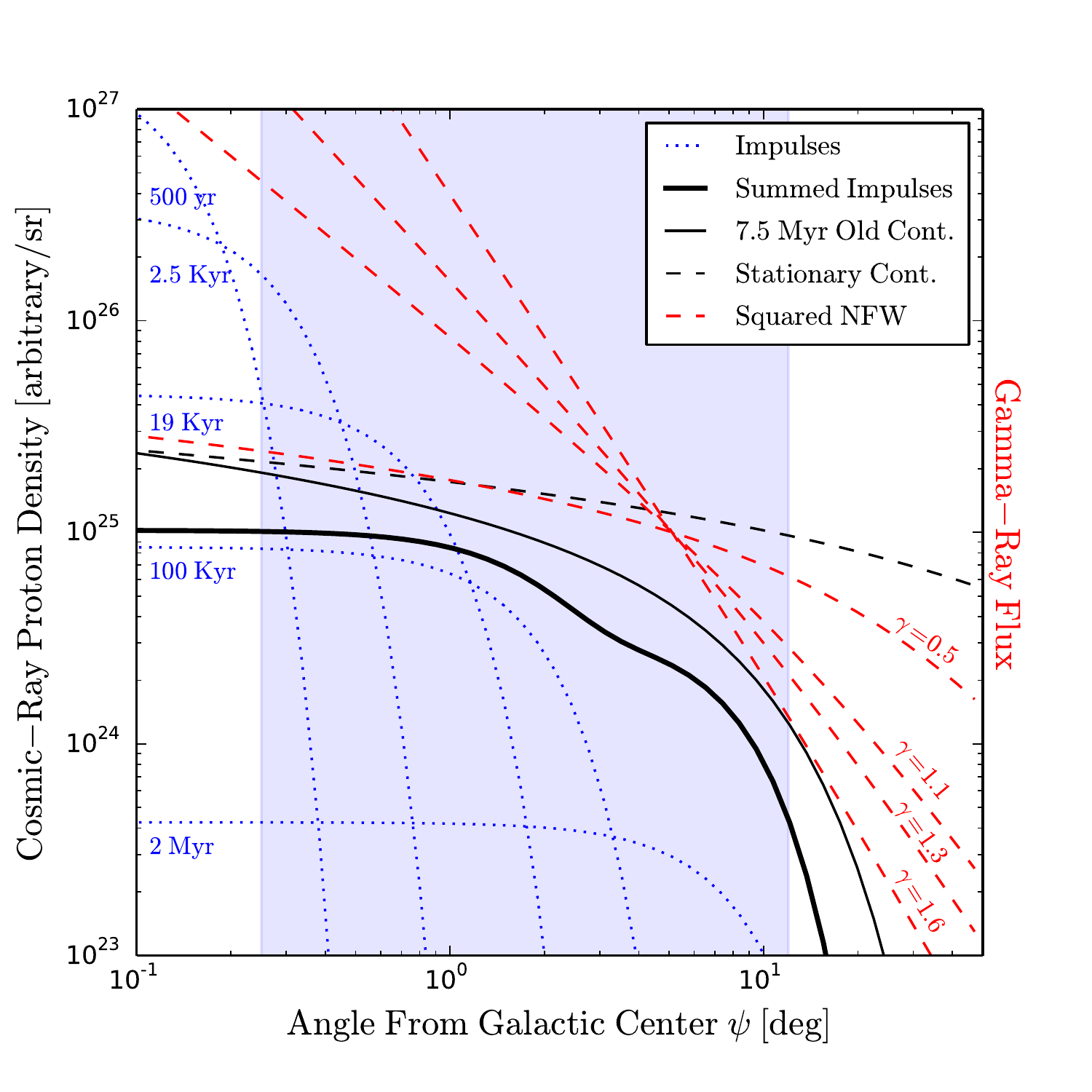}%
\end{centering}
\caption{\small The density of  cosmic-ray protons, at an energy of 2~GeV, projected along the line of sight as a function of the angular distance from the Galactic center in the spherically symmetric analytic diffusion approximation.  Shown in dotted blue lines is the evolution of an impulsive source after .5, 2.5, 19, 100, and 2000 Kyr from top to bottom.  We also show our summed impulse model (thick black), a 7.5 Myr old continuously emitting source (thin black), and a stationary continuous source (black dashed).  After a convolution with the gas density profile, the summed impulse and 7.5 Myr old continuous models have $\gamma$-ray flux profiles which approximately match that of an annihilating dark matter candidate following an NFW of inner slope $\gamma=1.3$ (shown in dashed red for several values of $\gamma$).  The shaded region shows the angular scales which are both above the Fermi-LAT point-spread function (lower-bound $\approx 0.25^\circ$) and bright enough to be differentiated from the background (upper bound $\approx 10^\circ-15^\circ$).}
\label{fig:morpho}
\end{figure}

 In Figure \ref{fig:morpho} we show the projected density of cosmic-ray protons for the putative impulsive and continuous sources listed in Table \ref{tab:sources}.  In particular, we show the evolution of a single impulsive source over the times from Table~\ref{tab:sources} as well as the continuous models C1 \& C2 along with a representative superposition of impulsive sources ($\rm{Im4}+10\times\rm{Im5}$) which we will employ in what follows. The overall normalization is left arbitrary for the sake of illustration.  It is crucial to note that this is the \emph{cosmic-ray proton} density, and that it must be multiplied with the spatially varying target gas density in order to obtain spatial distribution of the $\gamma$-ray flux. As a guideline, we also show the prompt {\em $\gamma$-ray} flux for an annihilating dark matter candidate following an NFW profile of inner-slope $\gamma$ between 1.1 and 1.3 and scale-length $r_s=24$~kpc.  These two values bracket the signal morphology resulting from the analyses of Ref.~\cite{Abazajian:2014fta, Daylan:2014rsa}.  
 
As we will demonstrate in the next section, the $\rm{Im4}+10\times\rm{Im5}$ and the C1 models have the correct proton densities to reproduce the GCE after convolution with the gas profile and are reshaped to closely match the $\gamma=1.3$ profile\footnote{The other values of $\gamma$ shown will be used in a later discussion of dust template modulation.}.   In the plot, the shaded region indicates the angular region of interest, bounded at low angular scales ($\approx 0.25^\circ$) by the point-spread function of Fermi-LAT, and at the approximate angular scales ($\approx 12^\circ$) where statistical and systematic uncertainties currently render the excess invisible over backgrounds. It is important to note that the recent bursts (Im1, Im2, Im3 and Im4), or superposition thereof, provide highly concentrated populations of cosmic-ray protons in the Galactic center, possibly yielding a bright, centralized, and spherically symmetric $\gamma$-ray emission.

Note that the time-scales we employ in the present estimates are not accidental: for example, model Im5 is close to the age of the Fermi bubbles, as estimated e.g. by Ref.~\cite{Guo:2011eg} and Ref.~\cite{Yang:2012fy} to be around 1-3 Myr, while model Im4 is also close to another alternate age estimate for the bubbles, $4\times 10^5$ yr, obtained by Ref.~\cite{Su:2010qj}, as well as matching age estimates of $10^4-10^5$ yr for the supernovae remnant Sgr A East at the Galactic Center. Also notice that for the time-scales listed above we are never in the regime where $t_1\gg \tau_{\rm pp}$ with the exception of the stationary continuous source, where protons are replenished over the region of interest anyway.

\subsubsection{Numerical Simulations}
\label{sec:nummorph}

The hadronic $\gamma$-ray emission from $\pi^0$ decay traces both the density of cosmic-ray protons and the spatial distribution of the target interstellar gas. While the discussion above shows that with one or more burst injections, a variety of cosmic-ray density profiles can be obtained (including highly centrally concentrated ones), the present discussion must include the target density for hadronic inelastic processes.  We note again that the template analyses of Refs. \cite{Daylan:2014rsa,Abazajian:2014fta} are predicated on a uniform distribution of cosmic-ray protons, and therefore neglect any gradients introduced by sources and by a non-trivial cosmic-ray morphology in the region of interest such as those shown in Fig.~\ref{fig:morpho}. 

In order to simulate in detail the $\gamma$-ray emission from the region and to assess the role of the cosmic-ray distribution, we employ the code {\tt Galprop v54.1.2423}~\cite{galprop}\footnote{Available at \href{http://sourceforge.net/projects/galprop/}{http://sourceforge.net/projects/galprop/}} which provides a 3+1-dimensional numerical solution to cosmic-ray transport along with empirically calibrated semi-analytical models of atomic, molecular, \& ionized hydrogen (HI, HII, H$^+$) gas in the Galaxy, in addition to a sophisticated treatment of pion production and decay.

For simulations longer than 50 Kyr we employ a {\tt Galprop} simulation consisting of a $10 \times 10$~kpc box centered on the Galactic plane with the x-axis defined by the Sun-GC line.  The half-height along the z-direction is 4~kpc with a lattice spacing of 200~pc along each axis.  For shorter simulations, the box-size is reduced to a sufficiently large cube of dimension 4~kpc with lattice spacing reduced to 50~pc.  A source of cosmic-ray protons is then defined as a narrow sub-grid Gaussian localized at the Galactic center.  In the case of impulsive source models, the {\tt Galprop} code has been modified to inject protons in time following a $\delta$-function centered at $t=0$.  Cosmic-ray transport is then solved forward in time with the {\tt Galprop} code, using `explicit-time mode' with step sizes of $\Delta t=10^2, 10^3$ yr for sources younger and older than 50 Kyr, respectively.

As in the previous section, we assume an isotropic diffusion tensor with diagonal entries $D(E)=D_0(E/4~{\rm GeV})^{+0.33}$ and a diffusion constant $D_0=6.1 \times 10^{28}~{\rm cm^2 s^{-1}}$.  For our morphological study of impulsive sources, we have explicitly verified that the diffusion constant and the diffusion time (the ``age'' of the source) are approximately degenerate for the quantity $D_0 t_{\rm diff}$ held constant.  This is expected in the limiting case of Eq.~(\ref{eq:imp}) where the diffusion time is much shorter than the proton cooling timescale.  In other words, holding the quantity $D_0 t_{\rm diff}$ constant will preserve the shape of the diffusion cloud, although the flux scales as $D_0^{-1}$.  This implies that if the diffusion constant differs in the Galactic center our results will still hold, but diffusion timescales will change, as will the energetics in the case of a continuous source. 

The region of interest under consideration here extends to $\pm1.5~$kpc at 10 degrees, while the height of the diffusion zone is much larger and set to $\pm 4$~kpc.  Unless this half-height is reduced to $h_{\rm dif}\lesssim 2$~kpc, variations in the height of the diffusion zone are also of negligible impact, and are thus not considered. Diffusive reacceleration is incorporated using a Kolmogorov spectrum for interstellar turbulence ($\delta_{\rm turb}=1/3$) and an \alf velocity of 30 km/s. 

\rev{At the small Galactic latitudes of interest, low-speed ($\lesssim 15$ km/s) convective winds out of the Galactic disk have been confirmed to be negligible, via explicit simulations, and are set to zero.  Notably, recent studies \cite{Crocker2011a,Crocker2011b,Crocker2012,Yoast-Hull2014} have presented extensive multi-wavelength evidence for very fast ($\gtrsim 150$ km/s) global outflows from the Galactic center region.  Driven by intense and approximately constant star-formation, this energy independent advective transport provides a good fit over radio, GeV, and TeV observations and it is suggested that such a component could, in fact, dominate over diffusive transport.  A detailed model of outflows is beyond the scope of the present study, but should not alter our overall conclusions.  The narrow energy range of the GCE implies that diffusive transport is effectively energy-independent, and spherically symmetric advection should produce a comparable morphology in the inner galaxy, albeit with somewhat different time scales and energetics.  However, it should be kept in mind that at the level of morphological detail required for template analysis, such effects are important and could significantly change the quality of fit compared to templates derived using diffusion alone.}

As mentioned at the beginning of this section, a crucial ingredient that a full cosmic-ray simulation allows us to test is the role of the interstellar gas distribution in predicting the morphology of the diffuse $\gamma$-ray emission.  In our simulations, the interstellar gas consists of three components: molecular, atomic, and ionized hydrogen.  In {\tt Galprop}, the first two components are modeled as independent, cylindrically symmetric distributions of seven galactocentric rings derived from surveys of HI \& CO, where the latter is used as a tracer of molecular hydrogen~\cite{moskalenko:2002}.  These surveys are then combined with distance information derived from the line-of-sight velocity distributions and Galactic rotation curves to assign a gas density to each ring as a function of height. Finally column densities from the analytic model are renormalized to agree with the survey gas column densities, breaking the cylindrical symmetry and reproducing the observed asymmetric gas structures.  Using this gas model, {\tt Galprop} propagates the cosmic-ray protons and convolves the resulting density with the gas model in order to produce a projected map of the $\gamma$-ray flux.  The smallest resolvable scales are thus ultimately limited by the gas map resolution.  In the case of HI and $H_2$, this amounts to an angular resolution of 0.5$^\circ$ and 0.25$^\circ$ respectively. Notably, the latter is approximately of the same characteristic size as the Fermi PSF above a few GeV.

Within the Galactic plane, the mass fraction of ionized hydrogen is only a few percent when compared with the other two components. For the sake of comparison with the `inner-Galaxy analysis' of Ref.~\cite{Daylan:2014rsa}, we focus on Galactic latitudes $|b|>1^\circ$ where the ionized Warm Interstellar Medium (WIM) makes up a significant portion of the diffuse $\gamma$-ray signal. In {\tt Galprop}, the WIM is based on the commonly used NE2001 model of Cordes \& Lazio~\cite{cordes1,cordes2} with scale-heights doubled to 2~kpc to ensure consistency with recent pulsar dispersion data as described in Gaensler et al 2008~\cite{Gaensler}. 

We emphasize that our gas model is nearly identical to that used to derive the hadronic component of the Fermi-LAT Collaboration's Galactic Diffuse Model, although the Fermi diffuse model also includes inverse Compton scattering and bremsstrahlung contributions from high-energy electrons, which are not of interest in testing possible issues with the hadronic component of the diffuse emission.  For a thorough description of the gas model we discussed above, see Ref.~\cite{fermi_diffuse} and enclosed references.  One difference of limited importance in our implementation of the scale-factor $X_{\rm CO}(R)$.  This parameter captures the ratio between the survey-derived integrated CO line intensity and the $H_2$ column density. In contrast to the fixed value used by the Fermi-LAT team, we choose this ratio to increase as a function of  Galactic radius, in accordance with the findings of Ref.~\cite{Galprop_x_co}.  As this function is nearly flat in the inner Galaxy, this change is not expected to play a significant role.  

The gas model and diffusion setup are now defined and we thus proceed to a morphological comparison between centralized proton sources and the measured Galactic center excess.
In the analyses of Refs.~\cite{Abazajian:2014fta,Daylan:2014rsa}, the basic features of the excess emission show an approximately spherical shape with flux approximately 3\% of the brightness of the Fermi diffuse model in the central $5^\circ \times 5^\circ$ window~\cite{Abazajian:2014fta} centered on the GC.  We define three benchmark cases of interest: 

(i) a continuously emitting central source of high-energy cosmic-ray protons, which has reached steady state over $\gtrsim 10^9$ year timescales, 

(ii) a continuous source which was started injecting protons 7.5 Myr ago, a time-scale consistent with ages proposed for the Fermi-bubbles, and 

(iii) a two-component impulsive source where protons were injected at ages of 19 Kyr, 100 Kyr, and 2 Myr, summed with free relative normalizations.  

In what follows, we calculate the $\gamma$-ray emission profile of our models as a function of the projected distance from the Galactic center.  We then fit this profile to the GCE to determine statistical compatibility and study the remaining spatial properties.

\begin{figure}[t]
\begin{centering}
	\includegraphics[width=\columnwidth]{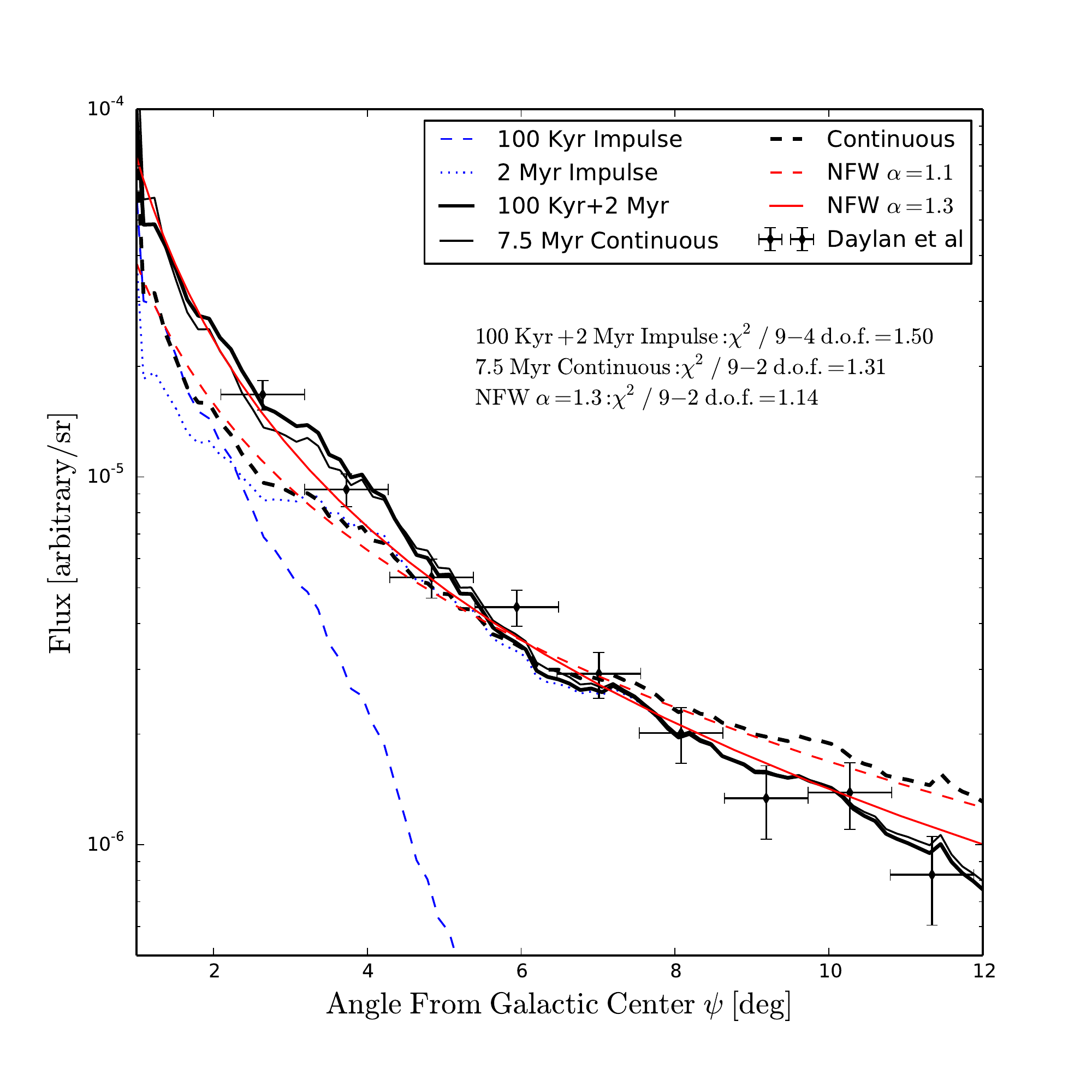}
	\end{centering}
\caption{\small Projected flux density at 2~GeV as a function of from a proton source at the Galactic center. For non-dark matter lines, results are derived from a full {\tt Galprop} simulation of diffusion and subsequent neutral pion decay averaged over the north + south regions with the Galactic plane ($|b|\pm 1^\circ$) masked out upon integration. In black we show radial flux profiles for our summed impulsive (thick), a 7.5 Myr old continuous source (thin), and a steady-state continuous source (dashed).  In blue-dashed and blue-dotted we show the individual impulsive sources at 100 Kyr and 2 Myr.  Finally, we show NFW profiles with inner slopes 1.3 and 1.1 in solid and dashed red. Data points are taken from Daylan et al (2014)~\cite{Daylan:2014rsa}.}
\label{fig:flux}
\end{figure}

In Figure~\ref{fig:flux} we plot the projected $\gamma$-ray flux, integrated along the line-of-sight and assuming a solar position of $r_\odot=$8.5~kpc, for each model as a function of radius from the Galactic center and compare against the `concentric ring' analysis of Daylan et al (2014)~\cite{Daylan:2014rsa} (black data-points). In practice, we use the same convention for this figure as in Ref.~ \cite{Daylan:2014rsa}: specifically, we average the line-of-sight integrated flux over circular annuli of increasing radius and a full-width of 1 degree, with the masked Galactic plane regions excluded.  Also shown for comparison are NFW profiles of inner slopes 1.1 and 1.3, as suggested in Ref.~ \cite{Daylan:2014rsa}.  In order to fit each model to the data, we choose normalizations using a (logarithmic) least-squares fit weighted by the (log) inverse variances of each of the nine points.  We then calculate the chi-squared per 9-2 degrees of freedom.  For the NFW models fit in Ref. \cite{Daylan:2014rsa}, the normalization and slope were free parameters.  For our proton source models, the normalization is allowed to vary and source ages were chosen by hand to provide a reasonable fit.  Both of these parameters are included when counting degrees of freedom.  In case (iii), i.e. the summed impulsive model, we do not include the 19 Kyr component since its contribution is negligible outside of the masked region (although it could be important to match the Galactic center analysis in the central few degrees, as we will show below).  We then sacrifice an additional degree of freedom and allow the normalization of the 100 Kyr and 2 Myr components to float independently.  Thus the summed model includes 2 ages and 2 normalizations.  The energetics of the normalizations are assumed arbitrary at this point. We will explore how reasonable the resulting normalization values we infer actually are in section \ref{sec:SNR} where a concrete astrophysical scenario is discussed.

The profile slope of the continuous source in steady-state appears to be slightly too flat to match the observed emission and does not provide a particularly good fit to the data ($\chi^2/\rm{d.o.f.}=6.15$).  However, if this emission were initiated at an age of $\mathcal{O}($5-10) Myr, the corresponding diffusion radius would be approximately 10 degrees. In this case, the resulting emission profile is significantly steepened, providing a very good fit -- $\chi^2/\rm{d.o.f.}=1.31$ -- compared to the $\alpha=1.3$ NFW profile where $\chi^2/\rm{d.o.f.}=1.14$.  Our best fitting model is the 100~Kyr+2~Myr impulsive model with $\chi^2/\rm{d.o.f.}=1.50$.
We find that for the summed impulse model, the best-fit injection luminosities have relative normalization 1:10, the larger corresponding to an event at 2 Myr.  Although this precise ratio depends on the relative ages of the two components, this fact does indicate that two events with relatively comparable energetics provides good agreement with the observed excess and may indicate that events of similar nature and origin might have fueled the two cosmic-ray bursts needed to explain the observed morphology.

\begin{figure*}[t]
\begin{center}
\includegraphics[width=2.1\columnwidth]{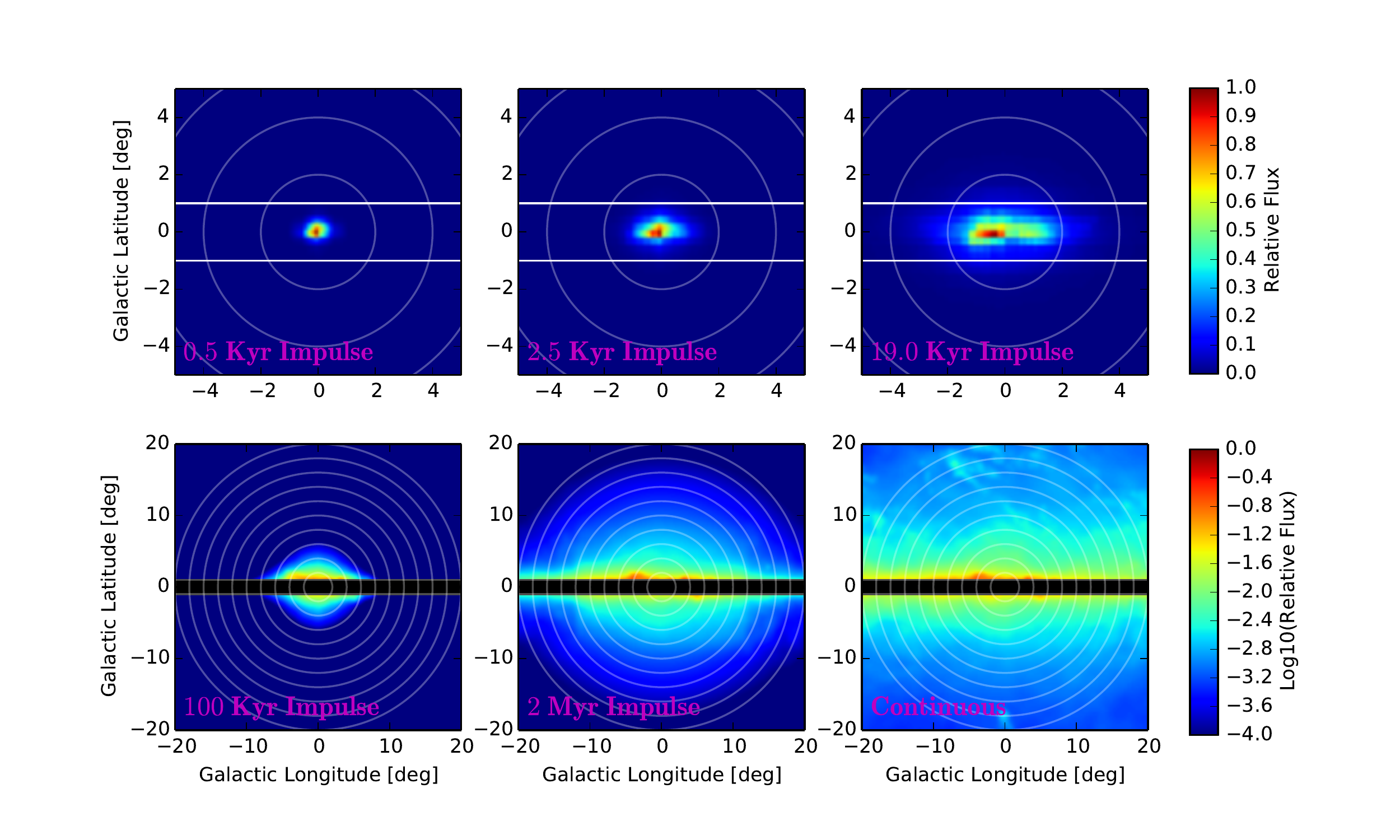}
\caption{\small Hadronic $\gamma$-ray flux density at 2 GeV from an approximately central source of high-energy protons integrated over the line-of-sight.  We show impulsive sources of increasing age in all panels with the exception of the bottom-right which shows a continuously emitting source in steady state.  For each map, the fluxes are normalized to the maximum.  For the ease of comparing the morphology of the claimed GCE in Ref.~\cite{Daylan:2014rsa} and shown in their fig.~9, we employ a linear scale in the three upper panels. The three lower panels employ, instead, a logarithmic scale to enhance the features of the emission outside the Galactic plane region. Also overlaid are reference reticles in increments of 2 degrees and indicators of the Galactic plane mask $|b|<1^\circ$. All maps have been smoothed by a Gaussian of width $\sigma=0.25^\circ$ to match Ref.~\citep{Daylan:2014rsa}.}

\label{fig:skymap}
\end{center}
\end{figure*}

In Figure \ref{fig:skymap} we investigate the overall spatial distribution of the emission from a new population of cosmic-ray protons injected in the Galactic center region.  The Figure shows the $\gamma$-ray flux associated with a central proton source for the benchmark impulse times of 0.5, 2.5 and 19 Kyr (upper panels) and of 100 Kyr, 2 Myr and continuous (lower panels). We use a linear scale in the three upper panels to help the Reader visually compare our results with what shown e.g. in Fig.~9, right panels, of Ref.~\cite{Daylan:2014rsa}. To the end of emphasizing the emission outside the Galactic plane, we instead employ a logarithmic scale for the older bursts and continuous sources in the lower panels. In each case, the fluxes are rescaled such that the maximum flux equals unity.  The Galactic plane mask ($|b|<1^\circ$) is bounded by white lines (or is masked out) and reference reticles have been overlaid at radial increments of 2$^\circ$.  

The top three panels show that a recent (from a fraction of a Kyr to tens of Kyr) impulsive cosmic-ray proton injection event in the Galactic center region yields a highly spherically symmetric and concentrated source, with morphological properties very closely resembling and matching those found in the Galactic center analysis of Ref.~\cite{Daylan:2014rsa} (see their Fig.~9, right panels), as well as in the GCE source residuals shown in the bottom panels of Fig.~1 in Ref.~\cite{Abazajian:2014fta}, and in the residual found in Ref.~\cite{gordon_macias:2013} and shown in Fig.~3. As long as the injection episode is recent enough, the morphology primarily traces the distribution of cosmic-ray protons, and is relatively insensitive to the details of the target gas density distribution --- the diametrically opposite regime from what assumed in the diffuse Galactic emission background models of Ref.~\cite{Daylan:2014rsa, Abazajian:2014fta, gordon_macias:201}.

It is evident that the sub-Myr simulations show a significant  degree of spherical symmetry outside  the masked regions.  Also, an excess with the same morphological aspect as in in fig.~9, right panels, of Ref.~\cite{Daylan:2014rsa} can be easily reproduced by young or very young sources, as shown in the three upper panels. As the diffusion time increases to to several Myr, the emission profile becomes more elongated and spherical symmetry is degraded. At higher latitudes ($|b|\gtrsim 2^\circ$), most of the spherical symmetry is, however, restored as the molecular and atomic gas distributions fall off, and the ionized component produces a more isotropic emission. In the template analyses of Refs.~\cite{Abazajian:2014fta,Daylan:2014rsa}, a portion of this residual ridge emission may also be absorbed by the Fermi diffuse model, although it is difficult to exactly pinpoint this effect without repeating the full maximum likelihood analysis.  It is also evident that gas structure is mostly washed out for recent impulsive sources, and that it becomes increasingly more prominent for older sources and for  the continuous emission cases.  Finally, we note that if a substantial portion of the inner excess is due to unresolved millisecond pulsars, much of the Galactic ridge would remain at a lower relative luminosity.

Quantitatively examining the angular profile for each source at a variety of different radii shows that within $\pm 45^\circ$ of the north and south Galactic poles, there is a high degree of spherical symmetry with typical (positive) variations on the order of 20\% with respect to the flux at Galactic north.  At larger angles, however, the flux rapidly rises as one approaches the Galactic plane to values many times larger than the Galactic north flux.  Although this does significantly illuminate the Galactic plane, it is unclear how important a role this plays in the analysis of Daylan et al \cite{Daylan:2014rsa}, where spherical symmetry was tested by scanning the axis ratio of the (now ellipsoidal) dark matter template.  Their analysis found a strong statistical preference in both the inner Galaxy and Galactic center analyses for an axis ratio of approximately $1:1\pm0.3$.  While this template distortion does provide a simple test, its geometry is not physically motivated and does not correctly probe the bar+sphere shape expected from a central hadronic source.

In Appendix C of Ref.~\cite{Daylan:2014rsa}, the authors examine the excess in two regions: north/south, defined by angles within the 45$^\circ$ of the poles, and east/west, defined as the complementary region dominated by the Galactic disk.  While both regions exhibit an excess, the E/W template shows a significantly enhanced peak of the signal compared to a flatter N/S spectrum~\cite{Daylan:2014rsa}.  This seems to indicate that either the Fermi-bubbles template absorbs much of the excess N/S emission, or that the emission is, in fact, more extended along the disk, as is seen in our benchmark models with a central cosmic-ray proton source.  In further testing the axis-ratio, Ref.~\cite{Daylan:2014rsa}, again, uses ellipsoidal projections of the NFW emission, this time allowing the template to rotate (there is still no test for a rectilinear disk component), finding a small statistical preference for an axis ratio of 1 to 1.3-1.4 elongated at an angle of $\approx 35^\circ$ counter-clockwise from the Galactic disk.  It is possible that this component of the excess is in fact a component of an extended central molecular gas bulge, as advocated e.g. in Ref.~\cite{gas_model}, which is oriented at $\sim 14^\circ$ CCW and is not modeled by the cylindrically symmetric {\tt Galprop} gas model and that, as a result, is therefore not included in Fermi Diffuse Galactic template.

In Appendix 4 of Ref.~\cite{Daylan:2014rsa} the hypothesis of an excess proton density is tested by adding an additional template based on the Schlegel-Finkbeiner-Davis dust map~\cite{SFD}.  The gas-correlated dust map is then spatially modulated so that the resulting template is given by 
\begin{equation}
\label{eq:modulation}
\rm{Modulation} = SFD(\vec{r}) \times \frac{\int_{\rm l.o.s.} \rho_{\rm NFW}^2(\vec{r})}{g(\vec{r})}
\end{equation}
where the NFW profile's inner slope was scanned to maximally absorb the emission, preferring an inner slope $\gamma=1.1$.  The functional form for $g(|l|,|b|)$ was then assumed to be the product of a latitudinal linear $\times$ exponential function and a longitudinal Gaussian. This function was then fit over $|b|<45^\circ$ and $|l|<70^\circ$ to also maximally absorb residuals.  It was found that the modulated dust absorbed a significant component of the excess when an additional NFW template was omitted.  However, when the NFW template was included in the analysis, it absorbed nearly the entire excess and the modulated dust map appears uncorrelated with the excess.  It was concluded that gas-correlated emission does not provide a suitable description of the GCE.  We disagree with this conclusion for the following reasons:  

\begin{enumerate}

\item The morphology of the underlying population of cosmic-ray protons which reproduces the GCE is shown by the 7.5 Myr continuous source shown in Figure \ref{fig:morpho} and is clearly \emph{very} different from any of the NFW profiles shown.  In the modulated dust template analysis, the functional forms chosen for $g(\vec{r})$ would need to be drastically different in order to reproduce distribution of protons matching that of Figure~\ref{fig:morpho}.  In particular, any analysis must consider that the target gas density already falls off as one moves away from the Galactic center, and that the dust map should be initially modulated by the expected proton density, \emph{not} proportionally to a projected NFW profile.  For example, if one takes $g(r)=1$ in Eq.~\ref{eq:modulation}, the resulting $\gamma$-ray template would fall off much faster than $r^{-3}$ when integrating over unmasked regions as was done for Fig.~\ref{fig:flux}.  As can be seen in Fig.~\ref{fig:morpho}, within the inner few degrees of the Galactic center, our 7.5 Myr continuous hadronic source would correspond approximately to a dust profile modulated by an NFW profile of inner slope $\gamma\approx 0.45$, which would then be required to steepen to more than $\gamma=1.6$ by $10^\circ$ in order to not severely overestimate the flux at large radii.

\item As seen in Figure~\ref{fig:skymap}, the gas-correlated emission from cosmic-ray populations younger than a few hundred Kyr remains highly spherically symmetric at high latitudes. Only in substantially older sources ($\gtrsim 1~\rm{Myr}$) does the gas structure of the Galaxy become completely dominant in shaping the $\gamma$-ray morphology.  In particular, this indicates that much of the dust structure lies at radii intermediate between the Earth and the Galactic center, whereas protons from a young cosmic-ray source have only reached the inner-most rings.  Our simulations take this 3-dimensional structure into account using gas velocity measurements to construct a model of Galactic structure and indicate that a 2-dimensional map of the column-density simply cannot account for a non-
uniform cosmic-ray density.

\item If astrophysical in nature, the residual is likely to be the result of several emission sources.  A substantial emission component in the inner few degrees naturally needs to be attributed to unresolved MSPs \cite{Yuan:2014rca} which exhibit an approximately spherically symmetric, or slightly ellipsoidal profile (see however \cite{Hooper:2013nhl}).  Such an addition would inevitably alter the preferred templates in the unmasked Galactic center analysis.

\end{enumerate}

To summarize this section, we have used the cosmic-ray propagation code {\tt Galprop} to simulate the $\gamma$-ray emission associated with neutral pion decay as cosmic-ray protons from a central proton source diffuse and interact with interstellar gas. Using a gas model identical to that of the Fermi-LAT Galactic diffuse template, we studied a variety of continuous and impulsive proton injection histories.  Under standard assumptions for the diffusion setup, it was shown that one can reasonably reproduce the spatial morphology of the observed Galactic center excess using source histories that are potentially correlated with past Galactic activity.  Specifically, the radial flux profile can be very closely matched if a continuous proton source turned on within the past 5-10 Myr, or if two or more events of comparable energy occurred at ages of around 0.1 and 2 Myr, although these simple benchmarks only represent a few possible scenarios.  The spatial distribution of these source's $\gamma$-ray emission may be somewhat more extended along the Galactic plane compared to the observed GCE, although without repeating the full likelihood analysis, a direct comparison is difficult. Indeed, a repeated likelihood analysis using the hadronic templates derived here is key to helping rule out a hadronic origin for the GCE and will be studied in detail in follow-up work.  The spatially concentrated excess found in the `Galactic center' analysis of Ref.~\cite{Daylan:2014rsa} is reproduced by young impulsive sources active from a fraction to a few Kyr ago in the center of the Galaxy, or perhaps even by efficient trapping of the 100 Kyr cosmic-ray population in sub-resolution molecular clouds at the GC. At Galactic latitudes above 2-3 degrees emission from the Galactic ridge becomes no longer dominant and at angles within $\approx \pm 45^\circ$ of the Galactic poles, our sources exhibit a very high degree of spherical symmetry while the projected gas structure is left largely unresolved relative to the steady-state Galactic diffuse model. Finally, we discussed possible correlations of the GCE with unmodeled gas components in the Galactic center as well as pointing out important issues with the modulated dust template analysis in Ref.~\cite{Daylan:2014rsa}.  In the next sections we turn to a study of the spectral characteristics of the GCE.

\subsection{Spectral Properties}
\label{sec:spectrum}

Three independent recent analyses of the GCE have found spectra which share a characteristic peak near 2 GeV, with little excess emission over background either below a few hundred MeV or above $10$~GeV. Although the location of the spectral peak is relatively robust, the shape of the excess is very sensitive to the modeling of point sources in the field, with additional systematic uncertainties such as the Galactic diffuse emission, and with differing ``regions of interest'', leading to a large variation in the reported low and high energy spectral slopes.  While most models are relatively well fit by a hard exponentially cut-off power law for the {\em photon} spectrum (and, as a result, reasonably well fit by dark matter models), we show below that a power-law {\em proton} spectrum with a break at energies of $\approx10$~GeV also provides good fits to the excess spectrum.  

A crucial feature of the differential $\gamma$-ray spectrum produced through the inelastic scattering of astrophysical high-energy protons on interstellar gas, is a characteristic maximum flux at 100 MeV induced by the rapid downturn of the inclusive $\pi^0$ production cross-section below 1~GeV.  Importantly, in the spectral energy distribution representation, $E_\gamma^2 dN/dE_{\gamma}$, this peak is shifted to $\approx 1 \rm{GeV}$ where both pulsar spectra and the GCE approximately peak.  It is thus a remarkable and unfortunate coincidence that the claimed GCE spectrally peaks at $\approx 2 \rm{GeV}$, where the likelihood of confusion with astrophysical sources is maximal. 

Here, we consider three reference spectral models for the underlying cosmic-ray proton population, and thus for the resulting $\gamma$-ray spectrum. The first cosmic-ray spectrum we consider is a power-law with an exponential cutoff (PLExp) where the proton spectrum at momentum $p_{\rm p}$ is given by,

\begin{equation}
n_{\rm p}(p_{\rm p})\sim\ p_{\rm p}^{-\Gamma}\ \exp[-p_{\rm p}/p_{\rm c}].
\label{eqn:plexpcut}
\end{equation}

The second and third models have a broken power-law injection of protons of the following functional form:

\begin{equation}
n_{\rm p}(p_{\rm p})\sim\ \left\{
     \begin{array}{lr}
	(p_{\rm p}/p_{\rm br})^{-\Gamma_1} & : p_{\rm p}<p_{\rm br}\\
	(p_{\rm p}/p_{\rm br})^{-\Gamma_2} & : p_{\rm p}>p_{\rm br},
     \end{array}
   \right. 
\label{eqn:BPL}
\end{equation}
where we allow the second index to be arbitrary in one case (BPL), and where we fix it to $\Gamma_2=\Gamma_1+1$ in the other (BPLFix).  The BPLFix model will later be motivated by the possibility of proton acceleration by supernova remnants taking place inside dense and partially ionized molecular clouds.  We then calculate the resulting $\gamma$-ray spectrum using one of {\tt Galprop's} newest models, employing the detailed low-energy parameterizations of Dermer (1986)~\cite{dermer:1986} with interpolation to the Monte-Carlo studies of Kachelrie\ss~\& Ostapchenko (2013) which better fit available collider data at high energies~\cite{Kachelriess:2012} (see App.~\ref{app:pion_decay} for details).

\begin{figure*}[t]
\includegraphics[width=2.1\columnwidth]{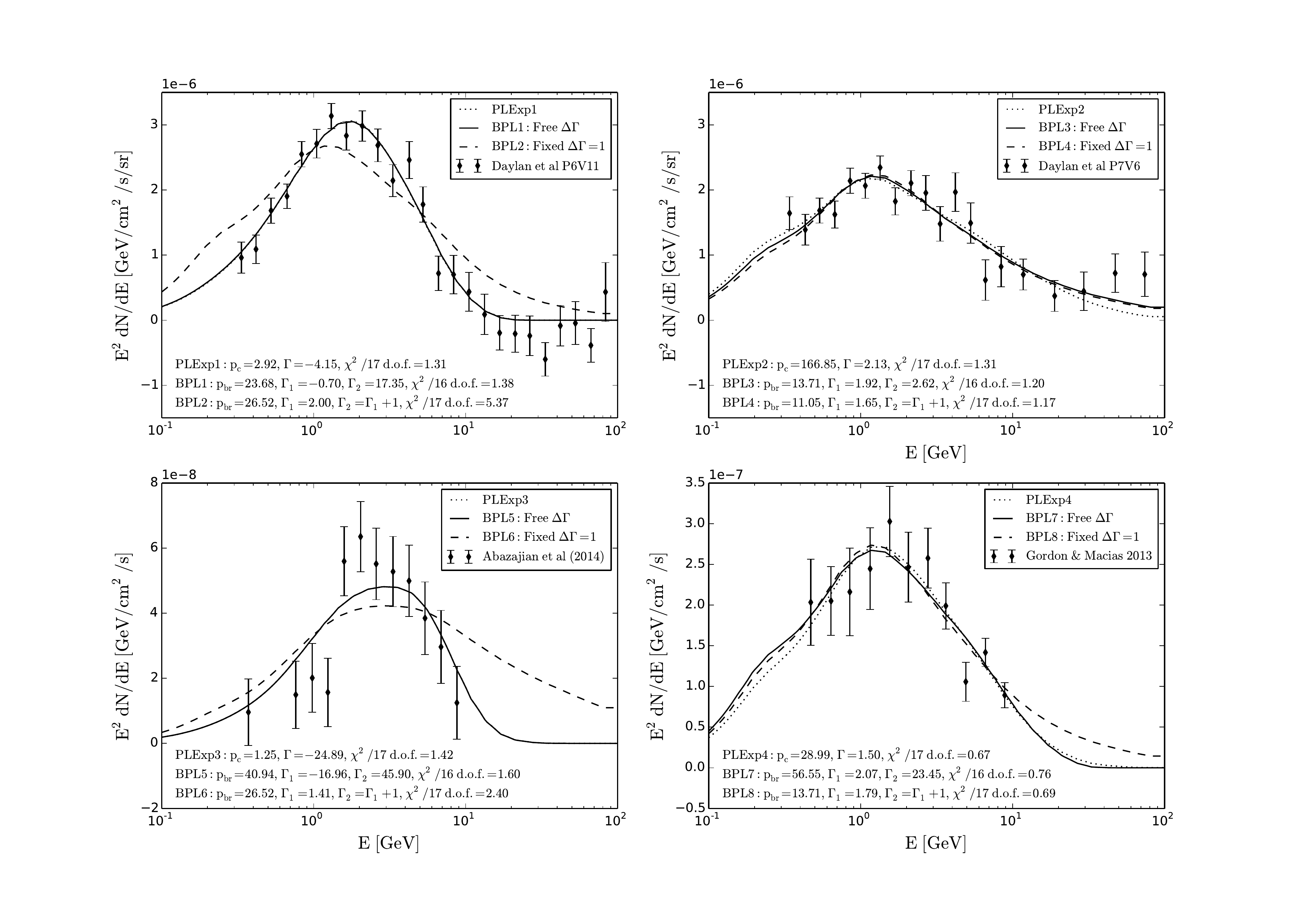}%
\caption{\small Best-fit $\gamma$-ray spectra for various analyses for the excess emission in the Galactic center region. In each panel we show three models of the underlying proton spectrum: Solid lines show the hadronic $\gamma$-ray emission for a broken power law proton injection spectrum where both indices and the energy of the spectral break are varied.  Dot-dashed lines employ the same functional form, but with the break in the spectral index fixed to $\Delta \Gamma=1$.  The dotted lines represent an exponentially cutoff proton spectrum.  In clockwise order and from the top left, the panels show data from Daylan et al Pass 6V11~\cite{Daylan:2014rsa}, Daylan et al Pass 7V6~\cite{Daylan:2014rsa}, Gordan \& Mac{\'{\i}}as~\cite{gordon_macias:2013}, and Abazajian et al \cite{Abazajian:2014fta}.  Note that the top row is normalized by the solid angle, while the bottom rows are integrated over the respective regions of interest.}
\label{fig:galprop_spectra}%
\end{figure*}

We then take data from the two analyses of Daylan et al (2014)~\cite{Daylan:2014rsa}, where different versions of the Fermi-LAT Galactic Diffuse Model were used to extract the GCE spectrum (using the template from the {\tt P6v11} and {\tt P7v6} releases, respectively), from Abazajian et al (2014)~\cite{Abazajian:2014fta}, and from Gordan and Mac{\'{\i}}as (2013)~\cite{gordon_macias:2013}.  We perform a maximum likelihood fit for each of our three spectral models, and compute the reduced $\chi^2$ for $f=N-M$ degrees of freedom where $N$ is the number of data points and $M$=3 for PLExp and BPLFix and 4 for the general BPL.  

It is crucial to note that Daylan et al \cite{Daylan:2014rsa} do not provide an estimate of the systematic uncertainties (which are expected to be relatively large), nor do we attempt to include any such estimate.  The error bars quoted in the analysis of Ref.~\cite{Daylan:2014rsa} arise purely from counting statistics.  Abazajian et al ~\cite{Abazajian:2014fta} do estimate the relative systematic error Galactic diffuse model based on variations in the spectral form chosen for the GCE, but they do not provide a specific number.  Based on their Fig.~8, we estimate the error (conservatively small) as $1\times 10^{-8}$ GeV/cm$^{2}$/s, and combine this in quadrature with the statistical errors for each point.  Gordan \& Mac{\'{\i}}as \cite{gordon_macias:2013} provide the most rigorous test of systematic uncertainties related to the Galactic diffuse model by looking at residuals as a window is scanned along the Galactic plane in regions with no contaminating point sources.  This results in an estimated $\approx 11\%$ standard deviation from Fermi's diffuse background model.  However, if we combine their statistical and systematic errors in quadrature, the fit is very poorly constrained. We therefore use only systematic uncertainties (which are typically larger) for this case. Below we discuss the results of Figure \ref{fig:galprop_spectra}, but one can already see from the substantial variations between the four extracted spectra that estimating the systematic  uncertainties is a highly non-trivial issue. We thus urge caution when interpreting the reduced $\chi^2$ values we quote, which should be taken only as a rough indicator of fit quality.

Figure \ref{fig:galprop_spectra}  shows the best fits for each of the three spectral models. In the top-left panel is the Daylan et al analysis which uses the the non-reprocessed {\tt P6v11} diffuse model. The excess is very well fit by the PLExp model which closely matches the prompt emission from a light dark matter candidate.  The BPL model also provides an exceptionally good fit, although the pre-break index is unphysically steep, at $\Gamma_1=-0.7$ while the second index converges to a value $\Gamma_2\approx$17 with a relatively large break energy $E_{\rm br}=23.7$~GeV, effectively mimicking the PLExp model (the two lines are in fact hardly distinguishable in the figure).  Of more interest is our BPLFix model, which provides a reasonable, though not optimal, fit to the data considering the underestimated error bars.  The best-fit low-energy index $\Gamma_1$=2 is intriguingly equal to the canonical value $\Gamma \approx 2$ expected from the theory of linear diffusive shock acceleration (DSA) thought to drive supernovae and black-hole acceleration processes.  \rev{Note that there exist systematic uncertainties arising in the low and high-energy ranges from modeling of the inclusive $pp\to\pi^0+$ anything cross section, as is discussed in App.~\ref{app:pion_decay} and Ref.~\cite{Stecker1973a}. Such uncertainties can be as large as 15\% below 1 GeV up to 40\% above 100 GeV, and thus affect any conclusion of the precise values needed for the cosmic-ray proton injection spectrum.}

The top-right panel shows the Daylan et al {\tt P7v6} analysis, which includes a Fermi-LAT model of the bubbles in the Galactic diffuse template in addition to the independent Finkbeiner bubble template.  Unlike the {\tt P6v11} analysis, which used mismatched photon data from the {\tt P7} release, this model is appropriately calibrated to the full {\tt P7} event data.  Compared to the {\tt P6v11} analysis, this approach yields a substantial flattening of the spectrum, with all models providing equally good fits, with nearly identical $\gamma$-ray spectra. $\Gamma_1$ is found to vary between 1.65 and 2.13 and in both BPL models $\Gamma_2\approx 2.6$.  The similarity between the BPL and BPLFix models is remarkable, given the significant difference in their initial spectral index.  This indicates a weak spectral dependence on $\Gamma_1$ due to the natural `GeV-bump' associated with pion decay.  This is also observed for in the fits to the other analyses, where the initial index can have completely unphysical values $\gamma_1\gtrsim 15$ with only a very small change in the log-likelihood.  Later we will show contour plots for the BPLFix model which indicate a strong covariance between the break momentum and the low-energy spectral index, and acceptable values of $\Gamma_1$ over the large range 1.25-2.5.

In the bottom-left panel we show spectra taken from the full model of Abazajian et al (Figure 3), Ref.~\cite{Abazajian:2014fta}, with statistical errors added as discussed above.  Even our conservative estimate of the systematic error leads to large uncertainties in the spectrum, and all of our models provide acceptable fits.  Although the BPLFix model does not appear to fit the data particularly well, we encourage the reader to review Figure 8 of Ref.~\cite{Abazajian:2014fta} where a range of GCE spectra are shown depending on the spectral model used in the likelihood fit.  The data shown here is for the measured residual -- as opposed to what results from a specific dark matter template -- and corresponds approximately to the most strongly peaked model.  The ``mean model'' of Fig.~8 in Ref.~\cite{Abazajian:2014fta} has a significantly softer low-energy spectrum.  The fit is also severely impacted by the asymmetrically small number of data points above the bump.  

Finally, in the lower-right panel we show data from Gordan \& Mac{\'{\i}}as (2013) which we found, again, to be well fit by all models, with a preference for a slightly hardened low-energy  index of $\Gamma_1$=1.73 for the BPLFix model and a break energy of 13.7 GeV.

Collectively, our results reveal two characteristic features: Firstly, in most cases there is a slight preference for the PLExp model; the BPL with free indices typically tend to converge towards a PLExp form.  One exception is the {\tt P7v6} fit from Daylan where the BPLFix model is actually preferred. The BPLFix models provide a reasonable fit throughout, with the exception of Daylan et al's {\tt P6v11} which, however, does not include any treatment of systematic uncertainties. Second, for a flat $p_{\rm p}^{-2}$ proton spectrum, the $\gamma$-radiation from $\pi^0$ decays naturally peaks at $\approx$1.25 GeV, slightly below the observed excess peak at $1.5-2$GeV.  In order to shift the peak to these higher energies we prefer a slightly harder initial spectral index $\Gamma_1$ between approximately 1.6 to 2, although there is low sensitivity to this parameter.  The placement of the spectral break is typically near $p_{\rm br}=10-50$~GeV and provides an effective control of the width of the spectral peak while the second index $\Gamma_2$ controls the cutoff rate as is expected from the nearly flat $\pi^0$ production cross-section above 1 GeV given in Eq.~(\ref{eq:pion_cross}).  The preference for a slightly hardened spectral index could arise naturally if the emission is a combination of e.g. SNR accelerated protons with index $\approx 2$ and MSP emission which can easily have Inverse-Compton spectra harder than 1.5. 

As an additional cautionary note, we reiterate that the theoretical predictions for the $\gamma$-ray spectra from proton-proton collisions are affected by significant systematic uncertainties associated to the modeling of the $pp\to\pi^0+$ anything production cross section. Such uncertainty feeds into the inferred spectral properties for the cosmic-ray populations associated with a given $\gamma$-ray emission. We discuss and evaluate quantitatively such uncertainties in the App.~\ref{app:pion_decay}. For now, it is important to note that any conclusion on the nature of the GCE based on spectral considerations alone ought to include this source of systematic uncertainty as well.

In addition to the `GeV bump' feature of the pion-decay spectrum, we point out the discussion of Section 4.2.3 in Ref.~\cite{aharonian}, which describes the temporal evolution of the spectrum of a cosmic-rays which are accelerated inside a molecular cloud, where large gas densities and magnetic fields can trap low-energy protons on timescales of $10^5$~yr.  For an impulsive accelerator and a cloud of very high density, high energy-protons can suffer substantial energy losses and propagate in a more rectilinear fashion, allowing escape while the low-energy protons remain inside.  The cloud is thus illuminated with a spectral energy distribution peaked at a few GeV with a steepened high-energy falloff at ages greater than $10^4$ years.  The low-energy index remains virtually unchanged unless the source is very young and brehmstrahlung from secondary electrons is contributing strongly.  By $10^5$~yr the cloud's peak flux decreases by 2 orders of magnitude and becomes part of the diffuse background.  Although this produces gas-correlated emission that could potentially be resolved, very close to the Galactic center the spatial resolution of Fermi-LAT is limited to scales larger than about 30 pc,   larger than most of the (many) molecular clumps orbiting in the central few parsecs. Such sources cannot thus be spatially differentiated from the central point source with $\gamma$-ray observations.  If the escaping high-energy emission is already suppressed, as in our BPLFix model, this would appear as an additional spectral break at approximately the same energy. This very scenario may be realized at the Galactic center for the $\sim 10^4-10^5$ year old supernova remnant, Sgr A East, which we discuss in detail later.  Almost certainly, molecular clouds are trapping protons at the Galactic center on scales unresolvable by Fermi-LAT and effectively reproducing the morphology of a younger source.

In summary, we proposed three models for the spectrum of a new population of cosmic-ray protons which could explain the GCE: an exponentially cutoff power law, and two broken power laws with free and fixed ($\Delta \Gamma=1$) changes to the spectral index, respectively. We calculated the $\gamma$-ray spectra resulting from inelastic collisions of the protons on interstellar gas, noting that nearly all physically reasonable proton injection spectra exhibit a bump near $\approx 1$~GeV in the $\gamma$-ray $E^2 {\rm d}N/{\rm d}E$ distribution.  For each model we performed a maximum likelihood fit to each of the four GCE residuals and found good fits in all cases over a broad range of parameter values.   We concluded that {\em the core spectral features of the GCE} -- namely a hard low-energy spectral index, a peak between 1-3~GeV, and a rapid decline above a few GeV -- {\em can be naturally produced by an additional population of cosmic-ray protons in the inner Galaxy}.  In the next section, we provide theoretical and phenomenological evidence that such a population is likely to exist in the Galactic center.

\section{Physical Models for the GC Excess}

In this section we demonstrate that the needed luminosity and spectral properties for the cosmic ray population we invoke to explain the GCE have sound physical motivations. In particular, we explain in Sec.~\ref{sec:breaks} how the spectral breaks in the cosmic-ray proton spectra we consider might have arisen in the Galactic center region, and related observational evidence; we then estimate in Sec.~\ref{sec:SNR} the energetics required by a cosmic-ray interpretation of the GCE, and argue that the time-scales and energy scales are plausible and in line with observations and theoretical expectations.

\subsection{A Mechanism and Evidence For GeV Spectral Breaks}\label{sec:breaks}
For half a century, the bulk of Galactic cosmic rays has been thought to originate from supernova remnants (SNRs) which inject 3-30\% of the total supernova energy ($\rm{E_{SN}} \approx 10^{51}$~erg) into protons and other light nuclei~\cite{pionSNR}.  A detailed theory of diffusive shock acceleration is still incomplete, but simplified linear models predict that supernova shocks propagating through an ionized gas precursor can accelerate protons and other nuclei up to $10^{15}$~eV with a resulting proton spectrum of $p_{\rm p}^{-2}$ at the source~\cite{blandford1987}.  When combined with sophisticated models of nuclear propagation through the Galaxy and solar system, this source spectrum successfully reproduces the locally measured spectrum of cosmic-ray nuclei. Direct confirmation of this acceleration model was provided only very recently (2013) by the Fermi-LAT collaboration following the detection of $\gamma$ radiation characteristic of $\pi^0$-decay in association with two known SNRs, IC443 and W44~\cite{pionSNR}.    

In order to postulate a viable astrophysical model for the Galactic center residual -- i.e. without invoking new particle physics -- we require either a substantial reduction in the $10^{15}$~eV high-energy cutoff, or a strong spectral break near $\approx$ 10 GeV which renders the signal invisible over that of the diffuse sea of background cosmic-rays where the $\gamma$ spectrum is roughly $\propto E_{\gamma}^{-2.7}$.  In what follows, we describe recent proposals that modify the canonical theory of DSA in the presence of dense molecular clouds which surround the inner Galaxy, as well as actual realizations of this scenario as seen in recent Fermi SNR observations showing significant breaks at $\mathcal{O}(10$~GeV) in the underlying proton spectrum.  It is thus possible to provide a natural explanation for the spectrum, energetics, and morphology of the GCE requiring only the assumption of an enhanced central supernova activity over the past few million years.

In DSA, shock waves propagating through ionized interstellar medium compress the plasma and transfer kinetic energy downstream through either two-body collisions, or through collective electromagnetic effects if the collision cross section is very small.  In the compressed zone preceding the shock front, resonant scattering of \alf waves efficiently accelerates particles until their gyro-radius $r_g=cp/(eB)$ exceeds the width of the shock layer~\cite{Drury1983}.  While this test particle case assumes a fully ionized cosmic-ray precursor, the Galactic center is only partially ionized, with well over 80\% of the gas content associated with neutral molecular hydrogen in the inner 200 pc, which completely engulfs the region of central starburst activity.  Malkov, Diamond, and Sagdeev \cite{Malkov2005,Malkov2011} demonstrated that when the upstream edge of supernovae shocks interact with molecular clouds, ion-neutral collisions effectively damp a range of  otherwise resonant \alf waves, severely deteriorating particle confinement within a slab of momentum space, and steepening the spectral index of protons by precisely one at an energy given in Ref.~\cite{Malkov2005} as
\begin{equation}
p_{\rm br}/m_{\rm p} c \approx 16 B_\mu^2 T_4^{-0.4}n_0^{-1}n_i^{-1/2}, 
\label{eqn:cutoff_energy}
\end{equation}
where $B_\mu$ is the magnetic field strength in units of $\mu G$, $T_4$ is the temperature of the ionized precursor in units of $10^4~K$, and $n_0$, $n_i$ are the neutral and ionized gas density given in in units of ${\rm cm^{-3}}$, respectively. Similar developments in non-linear DSA have shown that over 1-10 GeV the spectrum can be as steep as $E_{\rm p}^{-4}$ depending on the shock speed and environment, flattening out again above a few TeV~\cite{Blasi2012}.  

The mechanism described above successfully reproduces at least 6 of the 16 current Fermi-LAT observations of SNRs~\cite{fermi_snr1,fermi_snr2,fermi_snr3,fermi_snr4,Dermer2013c,fermi_snr5}, although the uncertainties associated with estimating the relevant environmental parameters are considerable. The 10 remaining observations have not yet incorporated this model into the analysis.  In Ref.~\cite{fermi_snr3}, several SNRs observed by Fermi were shown to be interacting with molecular clouds based on radio observations of 1720 MHz OH maser emission, providing a strong indication of shocked $H_2$.  The spectra were then reproduced by fitting the underlying proton distribution according to an exponentially cutoff power-law, as we do above.

SNRs interacting with highest density clouds were found to have low cutoff energies and hard proton spectra with [$\Gamma ,E_{\rm c}]=$ [1.7,160~GeV] and [1.7,80~GeV] compared to the low-density cases, where [2.4,1~TeV] and [2.45,1~TeV].  For another SNR, W44, an independent analysis found that the $\gamma$-ray emission was well fit by a hard proton spectrum of index between 1.74 and 2 with a cutoff at $p_{\rm c} \approx$ 10 GeV/c~\cite{fermi_snr2}. While these examples provide a representative sample of the expected range for the low-energy spectral index and cutoff energies, we do not necessarily expect a hardened spectrum to be correlated with high gas densities.  These SNR spectra match the $\gamma$ radiation expected from an exponentially cutoff proton spectrum quite well, possibly indicating that the theory of Ref.~\cite{Malkov2005} is underestimating the true breaking strength due to ion-neutral damping, or that an additional cutoff mechanism is at play.  In either scenario, a more pointed spectral peak is predicted, and as a result the fit to the residual GCE spectrum in Section \ref{sec:spectrum} is generally improved.

The Galactic center hosts a zoo of high-energy astrophysical sources including several SNRs, resolved \& unresolved pulsars, pulsar wind nebulae, and the central black hole Sgr A*.  Most notably Sgr A East is a $\sim 10^4-10^5$ year old and 10 pc wide SNR rapidly expanding into the molecular cloud M--0.02--0.07, where a half-dozen sites show also  show the 1720 MHz maser emission from shocked $H_2$~\cite{Yusef:1996}.  This complex encompasses the central black hole with most of the structure residing within a few parsecs from Sgr A* ($\lesssim 0.05^\circ$).  This separation is too small to be spatially resolved by Fermi-LAT, which has a maximal angular resolution of about a quarter degree, hence it will appear as a point source, perhaps with minor spatial extension, whose spectrum cannot be differentiated from additional Galactic center sources\footnote{For reference, the template analysis of Daylan et al, which uses large photon statistics and an event selection which optimizes PSF, finds the most likely position for the GCE to be centered within about 3 arcmin of Sgr A*. The next generation of ground-based $\gamma$-ray telescopes is likely to resolve these structures at energies above 50 GeV.}  

An especially intriguing candidate for the recent injection of cosmic-ray protons in the inner Galaxy is Sgr A East. As an estimate of the expected flux from Sgr A East, we utilize a similar object, SNR W44.  The latter is observed to have a differential flux of $\approx 1.25 \times 10^{-7}$ GeV/cm$^2$/s.  Multiplying by the square of the distance ratio $d^2_{\rm W44}/d^2_{\rm GC}\approx(2.9~\rm{kpc/8.3~kpc})^2$ we obtain a flux of $5 \times 10^{-8}$ GeV/cm$^2$/s, precisely in line with the GCE residual and the Sgr A* flux reported by Abazajian et al within a $1^\circ \times 1^\circ$ box centered on the GC~\cite{Abazajian:2014fta}. (Note that the the two Daylan et al fluxes reported in Figure \ref{fig:galprop_spectra} are normalized by the solid angle of a thin annulus at $5^\circ$ from the GC).  It remains to be assessed whether the spectral break energy near the Galactic center is compatible with the the results of Section \ref{sec:spectrum}, and whether a reasonable supernova rate is compatible with the observed flux.

The environment of Sgr A East has been studied in detail at radio and X-ray wavelengths.  Unfortunately, the complicated structure and rapid gradients in density, temperature, and magnetic field strength imply that there will be no single prediction for the spectral break energy predicted by Equation (\ref{eqn:cutoff_energy}), but, rather, a range of values dependent on the particular properties of the shocked region.  Here we expect that nearly all of the supernova activity will take place very close to the Galactic center, with conditions not far removed from those of Sgr A East.  The goal of the current study is to determine whether the conditions can plausibly reproduce the GCE, while a detailed environmental model and statistical treatment of uncertainties is reserved for future work.

\begin{figure*}[tb!]
	\begin{centering}
		\includegraphics[width=2.2\columnwidth]{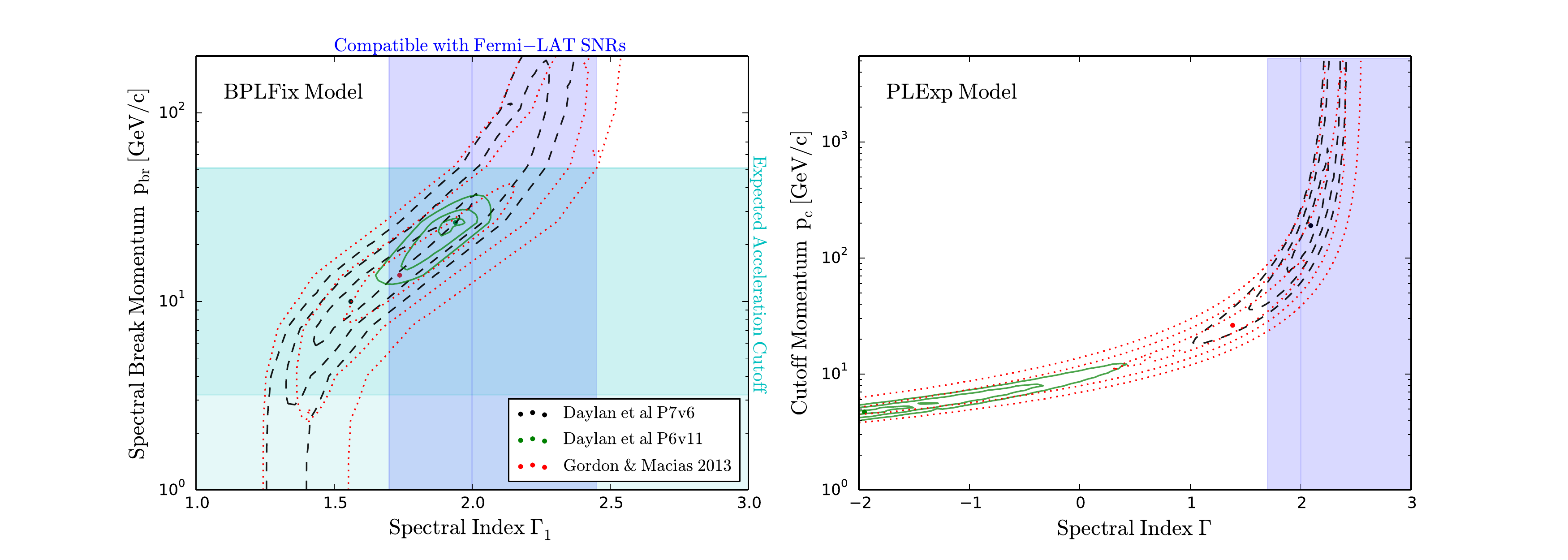}
	\end{centering}
\caption{ 1,2, and 3$\sigma$ confidence intervals for a broken power-law proton spectrum which steepens its index $\Gamma_1$ by one above the break energy $p_{\rm br}$ (left panel), and an exponentially cutoff power law (right panel), fit to three extractions of the Galactic center excess spectrum (excluding Abazajian et al). In the top panel, the bands shaded along the x-axis represent the range of low-energy spectral indices for SNRs interacting with dense molecular clouds as measured by Fermi-LAT in Ref.~\cite{fermi_snr3}. The dark and dark+light shaded bands along the y-axis indicate spectral break momenta expected to occur in dense molecular clouds and more ambient molecular densities respectively. Also note that confidence regions for the two Daylan et al spectra do not include any systematic errors and hence the true confidence contours are likely to be significantly more extended.}
\label{fig:contours}
\end{figure*}

The Central Molecular Zone (CMZ) is a large elliptical cloud with a gas mass fraction dominated by molecular hydrogen. It is thin and aligned with the Galactic disk, extending to a radius of approximately 150 pc from the Galactic center when projected along the line of sight\footnote{Interestingly, the same gas model in Ref.\cite{gas_model} finds a large gas bulge extending to 450 pc which is rotated 13.5$^\circ$ CCW from the Galactic plane when projected along the line of sight with an axis ratio of 3:1.  Daylan et al found a slightly preferred fit at roughly an angle of $35^\circ\pm$ CCW with an axis ratio of $1:1.4\pm .3$, possibly indicative of gas correlated emission.}.  This cloud makes up 5-10\% of the total Galactic molecular gas and is comprised of dense clumps of $H_2$ as well as of a lower density ambient component which completely fills the acceleration volume for any centralized SNR. In the inner 15~pc, typical densities can vary from the ambient value of $10^2~\rm{cm^{-3}}$ up to the dense molecular clouds at $10^5~\rm{cm}^{-3}$~\cite{Coil:2000,gas_model}, occasionally reaching even higher densities.  The warm ionized hydrogen is significantly more extended and provides the precursor for shock acceleration.  There is only weak power-law dependence of the break momentum on the density and temperature of the ionized component ($n_i^{-0.5}$ and $T^{-0.4}$).  Both of these components are reasonably well measured in the Sgr A* region using X-ray observations with ion densities near $10^3~\rm{cm^{-3}}$ and very hot plasma temperatures of $10^7$~K~\cite{Galactic_center_environment}.

The most important, and also the most uncertain factor in determining the break momentum, is the magnetic field strength in the shock propagation region. Zeeman splitting of OH molecules provides a measurement of the magnetic field strength along the line of sight, and indicates very strong fields in the large non-thermal radio filaments and possibly molecular clouds which can be as high as 1-4 mG~\cite{Yusef:1996,Ferriere2009} while Faraday rotation measurements indicate that the surrounding medium can be somewhat lower with a strength down to several hundred $\mu$G.  For an extensive review of magnetic fields in the Galactic center, we point the Reader to Ref.~\cite{Ferriere2009}.  

Efficient trapping of very low energy precursors in the very dense molecular clouds implies that these will be the primary acceleration sites for the resulting high energy cosmic-ray population, although a fraction will still originate from the surrounding lower density and lower magnetic field regions.  In this case, the lower densities of the ionized and molecular components partially cancel the effect of the smaller magnetic field on the break momentum, but some broadening of the spectral peak may be expected toward lower energies.   In order to estimate the range of break momenta achievable at the GC, we simply fix the least sensitive parameters to typical values, and set $n_i=10^3~\rm{cm^{-3}}$, $n_0=10^4~\rm{cm^{-3}}$, and $T=10^7$~K, while varying of $B$ between 0.5~mG and 4~mG. Doing this provides a break momentum between 0.79 and 51 GeV/c with a nominal value of 12.7 GeV/c for a 2~mG field strength.

Without more accurate measurements and high-resolution 3-dimensional models of the Galactic center environment, it is extremely difficult to definitively compute the resulting cosmic-ray spectrum.  If, in fact, these large magnetic fields are contained strictly to non-thermal radio filaments, or are much weaker then previously thought, as suggested in Ref.~\cite{YusefZadeh:2012nh}, the predicted momentum break would be significantly smaller, and the breaking mechanism would be disfavored as an explanation for the GCE.  It is also very likely that current conditions at the Galactic center differ substantially from those of 1-10 Myr ago especially if the Fermi bubbles formed on comparable timescales.  Compounded with uncertainties in non-linear DSA in the presence of ion-neutral damping, a conclusive statement is currently not possible.  Nonetheless, the observation of break energies from ten to several hundred GeV in nearby SNR indicates that such scenarios are not uncommon, and provide evidence that the description advocated above is not unrealistic.

In Figure \ref{fig:contours} we show confidence intervals for the low-energy spectral index and break energy for the BPLFix and PLExp models of the \emph{proton} spectrum as fitted to the two Daylan et al GCE residuals as well as that extracted by Gordon \& Mac{\'{\i}}as.  We do not show the results of the fits to the  Abazajian et al results due to the previously mentioned asymmetry in the number of points below and above the spectral peak which forces a very hard spectrum that clearly does not fit the rapid falloff above 2 GeV seen in the other datasets.  While the \emph{residual} found by Abazajian et al is indeed very hard at low energies, when an additional GCE template and spectral form is included as part of the fit, the low-energy index softens significantly becoming very similar to the other analyses.  This behavior is clearly delineated in Fig. 8 of Ref.~\cite{Abazajian:2014fta} and the enclosed discussion. 

In the left panel, the shaded regions along the x-axis show the range of the low-energy proton index which are compatible with Fermi-LAT observations of SNRs interacting with molecular clouds taken from Refs.~\cite{fermi_snr2,fermi_snr3}, highlighting the canonical index $\Gamma_1=2$ predicted by linear DSA.  In the shaded y-axis regions, we show expectations for the position of the spectral break in conditions typical of the very dense molecular clouds (dark cyan) and in the ambient lower density environment (darker+lighter cyan).  It is promising that these contours are fully compatible with one-another when fitting to the BPLFix model.  Clearly, if one assumes the BPLFix model, the parameter values are in line with those expected from SNR interacting with molecular clouds in the Galactic center. 

In the right panel we show similar regions shaded along the x-axis representing the range of the spectral indices compatible with Fermi-LAT observations where fitting the underlying proton spectrum used an exponentially cutoff power-law model~\cite{fermi_snr1,fermi_snr4,fermi_snr5}. Although these studies also indicate GeV-TeV scale cutoff energies, it is unclear how such cutoff scales should change in the Galactic center environment without a theoretical understanding of the cutoff mechanism itself.  In contrast to the BPLFix model, a PLExp spectrum reveals less compatibility among the three GCE residuals, with the main {\tt P6v11} analysis of Daylan et al requiring an unphysically hard spectral index.  Interestingly, two of the GCE datasets show a rapid upturn in the contour as the spectral index rises above $\Gamma=2$.  In this region, the fit is almost completely insensitive to the cutoff energy up to at least $\approx 10$~TeV. Notably, a spectral index softer than 2 is commonly invoked when modeling radio and $\gamma$-ray emission from AGN in the context of hadronic injection.  Although the relatively low momentum cutoff would still need to be explained, the insensitivity here could allow for a variety of possibilities, and warrants additional study.

To summarize, we find that the occurrence of a break in the spectrum of cosmic-ray protons in the specific environment of the Galactic center is well-motivated. Observations of the $\gamma$-ray spectrum of several SNR with the Fermi LAT point to cosmic-ray proton spectral features aligning precisely with those needed to fit the spectrum of the GCE; the location of a spectral break in the accelerated cosmic-ray protons in the presence of dense molecular clouds in the inner Galaxy also falls squarely in the range that optimally fits the inferred $\gamma$-ray spectrum of the GCE. We thus conclude that the spectra we invoked to fit the GCE are well motivated by both theory and observation.

\subsection{SNe Rates and Starburst Histories}
\label{sec:SNR}

In this section we explore the energetics required to produce the GCE with cosmic-ray protons injection at the center of the Galaxy.  In the previous section, we showed that the flux measured from SNR W44 corresponds to the approximate luminosity needed to explain the GCE in the inner Galaxy.  At radii larger than 1 degree, the GCE signal decays rapidly as shown in Fig.~\ref{fig:flux}.  In Section~\ref{fig:flux} we showed that such a radial flux profile could be achieved rather naturally by the diffusion of protons injected at the Galactic center in several different episodes -- for example, impulsive injection over 2-3 different epochs ($\approx 10^4, 10^5,\ \rm{and}\ 10^6$~yr) or continuously if the source was turned on around 7.5 Myr ago.  Previously, we ignored the normalization of the flux and were only concerned with the relative normalization of the summed impulsive models  This revealed that the 100 Kyr + 2 Myr summed model preferred relative normalizations of, respectively, 1:10. The energetics of these long-timescale events is more constrained than for more recent outbursts. 

We compute the $\gamma$-ray flux due to protons assuming a nuclear injection spectrum of index $\Gamma_1=2$ breaking to $\Gamma_2=3$ at 10 GeV.  We find that the $100$ K and $10^6$ summed impulsive model requires a total injection of $\mathcal{O}(10^{52})$~erg into protons with energies above 100 MeV in  order to produce flux compatible with the GCE consistent with the very recent findings of~\cite{Yoast-Hull2014}.   For continuous sources only a few million years old, the required energy is approximately $10^{38}$~erg/s, or a few $10^{48}$~erg/century, while continuous sources in steady-state are an order of magnitude less and comparable to the rates needed to maintain the current molecular gas temperatures near the Galactic center~\cite{Yusef-Zadeh2012}.  

Stellar densities at the Galactic center are extremely high rising from a mass density of $10^4$ $M_\odot/{\rm pc}^{-3}$ at a radius of 10 pc to over $10^6$~$M_\odot {\rm pc}^{-3}$ in the central parsec (compared to the local density $\ll 1~M_\odot/{\rm pc}^3$).  Measurements of the infrared luminosity near the Galactic center provide an indirect probe of the star formation rate. If this has not changed dramatically over short stellar evolution timescales ($10^8$~yr), the expected supernova rates are 0.01-0.1 per century~\cite{Crocker2010} each injecting $\epsilon_{p} 10^{51}$~erg where $\epsilon_p$ is the fraction of the supernova energy channeled into proton acceleration, often taken to be near $0.1$~\cite{aharonian}.  This implies an average continuous injection rate of $10^{48}-10^{49}$~erg/century, compatible with the observed excess signal.  For impulsive sources, the same value of $\epsilon_p$ would require bursts of 10-100 supernovae to occur within a timescale relatively short -- $10^4$ to $10^5$~yr -- with respect to the diffusion timescale.  While any realistic scenario would likely be an admixture of continuous and burst-like injections, the supernova rates required to reproduce the observed GCE flux in either case are well within the possible histories of the Galactic center Region.


\rev{Star formation rates within the central hundred parsecs of the Galaxy is a subject of hot debate.  Over $\sim$10~Gyr timescales, several studies\cite{Crocker2011a,Crocker2011b,Crocker2012} suggest that the star formation rate has been approximately stable, with long-lived bulge stars formed during the Milky Way's last major merger event and relatively quiescent activity since.  On much shorter timescales the situation is less clear. Highly variable and intense star-formation producing tens to thousands heavy stars over a few Myr, cannot be ruled out. High ionization rates, severe shocks, and the large scale inflow/outflow accompanying molecular cloud collisions or cataclysmic events, such as star-bursts or activity from the central supermassive black hole, can trigger periods of rapid star-formation taking place inside the densest molecular clouds\footnote{For a recent review of massive star formation in the Galactic center, see Ref.~\cite{Figer2008}}. In contrast to self-collapsing molecular clouds, such external compression mechanisms are believed to induce significantly heavier initial mass functions, producing O/B type stars which evolve over $10^6-10^7$ years before going supernova~\cite{1993ApJ...408..496M}.  While many of the Galactic center conditions can also inhibit star formation, observations indicate at least 100 high mass stars with ages estimated around several Myr, indicating that an era of high star-formation rates may have occurred $\sim 10^7$ yr ago which has since halted.}

It is notable that the orbital time period for a typical molecular cloud at a radius of 1 pc is $10^5$ years providing ample opportunity for interactions with other clouds, or with the accretion disk surrounding the central black hole~\cite{1993ApJ...408..496M}.  Alternatively, this could be taken as possible evidence of intense supernovae or Sgr A* activity several million years ago in which shocked molecular clouds became highly compressed, initiating star-formation.  Supernovae bursts have also been proposed as a driver of the Fermi bubbles on Gyr timescales~\cite{Crocker2010} and as a mechanism to explain the extremely hot plasma temperatures in the Galactic center where gas in excess of up to $10^8$~K are observed, hotter than the Galactic escape energy, implying extraordinary energy injection event(s) with total energy $10^{53}$~erg and a lifetime of order $10^5$ yr in order to remain contained near the Galactic center~\cite{Galactic_center_environment}.  Such extreme events have comparable timescales and energetics to produce the scenarios explored earlier. 

Another possibility which has been previously considered is the injection of protons directly from the central black hole~\cite{2011ApJ...726...60C,2012ApJ...753...41L,2011PhRvD..84l3005H}.  Our morphological analysis of Section~\ref{sec:morphology} is substantially blind to the spectrum over the very narrow energy range under consideration.  Spectrally, the situation is more difficult as such low-energy cutoffs in the proton spectrum do not seem typical of active galaxies\footnote{Although many AGN spectra do have breaks in the $\gamma$-ray spectrum near 5 GeV, this results from absorption in the so-called `broad-line region' within a few hundred Schwarzschild radii of the central black hole and does not provide a viable mechanism for \emph{extended} emission peaked in the GeV range.}.  It is possible that a yet unknown mechanism is responsible for producing a cutoff proton spectrum from Sgr A*.  Such a scenario was in fact considered in Ref.~\cite{2005ApJ...619..306A}.  In this case, the black hole is taken to be in a quiescent state with a very hard proton spectrum $\Gamma=1$ exponentially cut off at 5 GeV.  Secondary electrons produced in the hadronic interactions are of low enough energy to preserve their spectral shape and emit very hard infrared and millimeter synchrotron spectra, matching radio observations of Sgr. A*. In such a scenario, the soft protons could diffuse to large radii while the hard synchrotron emission would be confined to the confined to the ultra high magnetic fields in the immediate vicinity of the central black hole.

A very recent result~\cite{Yoast-Hull2014} examined the compatibility of radio and GeV/TeV $\gamma$-ray observations with predictions from two models of starburst galaxies based on the interactions of cosmic-rays (of supernova origin) in the Central Molecular Zone.  Particularly careful attention was paid to the relevant astrophysical parameters, which are fully consistent with what we employed here. For each model, the average magnetic field strength, convective wind speed, ionized gas density, and free electron absorption fraction were allowed to vary in order to find optimized fits to data.  The results strongly favor ion densities between 50 and 100~cm$^{-3}$ and magnetic fields between 100 and 350~$\mu$G throughout the \emph{entire} ambient CMZ cloud.  While radio and TeV observations are well fit, a GeV excess still persists.  The addition of additional populations of either protons or electrons is then considered. In the case of protons, a soft spectral index $\Gamma\approx 3.1$ and a supernova rate enhanced by a factor $\sim 100$ are found to be consistent, but are dismissed from further analysis based on the required SNe rate.  For electrons, the energetics are more compatible, but the spectral indices predicted for radio and $\gamma$-rays are found to be inconsistent with observations. We note that in our analysis, this is precisely the proton spectral index we predict above a $\sim 10$~GeV and that the required SNe rate is substantially reduced due to our much harder $\Gamma=2$ low-energy spectrum (which also matches the GeV excess in much greater detail than what considered in Ref.~\cite{Yoast-Hull2014}). We find it remarkable that a completely independent analysis of the conditions required to fit starburst models to observations of the CMZ can naturally motivate an SNR explanation for the GCE at such a detailed level.

To summarize, in this section we have presented observational and theoretical evidence for spectral breaks in the cosmic-ray spectrum when protons undergo diffusive shock acceleration by supernovae remnants which are inside or strongly interacting with partially-ionized molecular cloud complexes.  This ion-neutral damping mechanism then predicts a break in the power-law index $\Gamma$ of precisely $\Delta\Gamma=$1 occurring at an energy parametrized by the local magnetic fields, ion/neutral number densities, and the temperature of the ionized precursor.  We then discussed the conditions in the Galactic center environment needed to produce plausible break energies which were found to be of the correct order to explain the GCE if magnetic fields in the acceleration region are of approximately mG strength.  Allowing the spectral index and break energy to float, we presented confidence contours for our fit to the Galactic center excess and showed that the preferred parameter space is spectrally compatible with an interpretation in terms of protons originating from GC SNR.  Next, the energetics required to match the GCE were calculated, finding that each impulsive event requires tens to hundreds of supernovae (total energy $10^{52}$~erg) to occur on timescales somewhat smaller than the age of the outburst, or that quasi-continuous sources inject protons at a rate of order $10^{38}$~erg/s.  Finally, we discussed evidence for sporadic increases in star-formation and supernovae rates in the Galactic center on the timescales relevant explain the extension of the GCE in terms of cosmic-ray diffusion and subsequent $\gamma$-rays of hadronic origin.  Very large uncertainties plague each step of such an analysis and estimate.  Nonetheless, the combination of spectral compatibility along with reasonable energetics and a plausible Galactic history provide a crucial background to any analysis of $\sim$GeV $\gamma$-ray data at the Galactic center.


\section{Discussion and Conclusions}
We presented a case for high-energy cosmic-ray protons injected in the Galactic center region as a plausible explanation to the reported Galactic center $\gamma$-ray excess over the expected diffuse background.  Our study focused on whether such an explanation meets the required (i) morphology, (ii)  spectrum and (iii) energetics. 

We demonstrated that cosmic rays injected on the order of a mega-year ago explain the observed spherical symmetry reported from the ``inner Galaxy'' analysis of Ref.~\cite{Daylan:2014rsa}, while a more recent (on the order of a few kilo-years old) episode would possess the same morphology obtained for the innermost portions of the Galaxy in the ``Galactic center'' analysis of Ref.~\cite{Daylan:2014rsa}

We showed that the $\gamma$-ray spectrum predicted by cosmic-ray proton energy distributions responsible for the emission observed from supernova remnants (such as broken power laws with specific spectral indexes, and exponentially suppressed power laws) provide excellent fits to the observed Galactic center excess. We pointed out that the preferred range for the break of the power law and for the spectral indexes inferred from the observed excess fall squarely in the ranges inferred from observations of supernova remnants, as well as in the general range expected from theory considerations. We also pointed out the importance of systematic effects in spectral reconstruction due to hadronic cross sections impacting the predictions for the $\gamma$-ray spectrum from inelastic proton-proton collisions.

Finally, we inspected the time-scales, spectrum and energetics we invoked to reproduce the morphology and spectrum of the Galactic center excess in the context of one or more additional populations of cosmic-ray protons in the region. We demonstrated that the existence of such populations is motivated by a variety of observational and theoretical reasons, which we reviewed in detail.

In conclusion, with the present study we gave proof of existence of a well-motivated alternative to dark matter annihilation or milli-second pulsars as an explanation to the reported Galactic center $\gamma$-ray excess. Our results indicate that conclusively claiming a signal of New Physics from $\gamma$-ray observations of the inner regions of the Galaxy must contend with a variety of additional astrophysical processes. In particular, we highlighted that one or more previously unaccounted-for populations of cosmic-ray protons in the Galactic center could potentially produce a $\gamma$-ray emission with a spectrum, morphology and intensity closely resembling those of the Galactic center $\gamma$-ray excess.



\section*{Acknowledgments}
%
\noindent We thank Andy Strong, Amy Furniss, Roland Crocker, Gudlaugur J{\'o}hannesson, and Tim Linden for very helpful discussions. EC is supported by a NASA Graduate Research Fellowship under NASA NESSF Grant No. NNX13AO63H. SP is partly supported by the US Department of Energy, Contract DE-FG02-04ER41268. 
%

\begin{appendices}
\section{Uncertainties on $\pi^0$ Emissivities}
\label{app:pion_decay}

The $\gamma$-ray emissivity $q_\pi(E_\pi)$ of secondary neutral pions produced through inelastic scattering of cosmic-ray protons on interstellar hydrogen is given by the following expression:
\begin{equation}
q_\gamma(E_\gamma)=2\int_{E_{\rm min}}^\infty\frac{q_\pi(E_\pi)}{\sqrt{E_\pi^2-m_\pi^2}}{\rm d}E_\pi,
\end{equation}
 where $E_{\rm min}=E_\gamma+m_\pi^2/(4E_\gamma)$ and the neutral pion production term,$q_{\pi}$,  is defined by,

\begin{equation}
q_\pi(E_\pi)=4 \pi n_{\rm H} \int_{m_{\rm p}}^\infty j_{\rm p}\left(\sqrt{E_{\rm p}^2-m_{\rm p}^2}~\right) \frac{\rm{d}\sigma_{{\rm pH}\to \pi^0}(E_{\rm p}, E_\pi)}{{\rm d} E_\pi}~{\rm d}E_{\rm p},
\end{equation} 
with $n_{\rm H}$ the target hydrogen gas density, $\sigma_{{\rm pp}\to \pi^0}$ the inclusive $\pi^0$ production cross section (p + p$\to \pi^0 +$ anything), and $j_{\rm p}(p_{\rm p})$ the cosmic-ray proton density as a function of the proton momentum, following recent results from Ref~\cite{Dermer2012}.  Note that many references use instead a proton spectrum following $E_{\rm tot}$ rather than $p_{\rm p}$ or kinetic energy $T_{\rm p}$.  Although these asymptote to each other at $E\gg m_{\rm p}$, the assumption can have a non-negligible impact on the low-energy $\gamma$-ray spectrum for soft \emph{proton} spectra $\Gamma \gtrsim 2.5$, where the low-energy protons contribute heavily.  Since the cross-section falls off very rapidly below 1 GeV, this is negligible for the harder spectra of interest here.  Remarkably, this cross-section is still not known to better than $\pm$10-20\% near the pion production threshold of $T_{\rm p}=280$~MeV up to a few GeV, resulting in an important systematic uncertainty when using the $\gamma$-ray spectra to probe the underlying spectrum of nuclear cosmic-rays, or {\it vice versa} as is the case here. Until improved laboratory measurements are made available this remains a limiting factor in determining the global spectrum of diffuse Galactic protons using Fermi-LAT photon data~\cite{Dermer2013,Dermer2013a}.  In this Appendix we demonstrate the systematic variations between four common models of the pion emissivity. 

The first model we consider is the simple $\delta$-function approximation for the cross section\cite{aharonian} as parametrized in Ref.~\cite{Gaisser1990}; we then consider the three numerical models implemented in {\tt Galprop}, which use cross-sections from Kamae et al (2006)~\cite{Kamae2006}, Dermer (1986)~\cite{dermer:1986}, and the model used throughout this paper: a combination of Dermer (1986) near threshold and interpolated to Kachelrie\ss~\& Ostapchenko (2013) at higher energies~\cite{Kachelriess:2012}, hereafter DKO.

\rev{The simplest estimate of the pion emissivity is obtained in the delta-function approximation, where proton-proton collisions are assumed to produce only pions and hence the well known inelastic cross-section is used as a proxy for the inclusive $\pi^0$ cross section and}

\begin{align}
 q_\pi(E_\pi) &= \frac{n_{\rm H}}{\kappa_\pi}\sigma_{\rm pp}^{\rm inel}\left(m_p+\frac{E_\pi}{\kappa_\pi}\right)\\
  &\times j_{\rm p}\left( \sqrt{\left(m_p+\frac{E_\pi}{\kappa_\pi}\right)^2-m_{\rm p}^2}\right),
\end{align} 
\rev{with cosmic-ray proton density $j_{\rm p}$, and with $\kappa_\pi\approx0.17$ the mean fraction of the impinging proton kinetic energy transferred to the secondary $\pi^0$ per collision \cite{Gaisser1990}.  This has been adjusted empirically to provide better excellent agreement with Monte Carlo simulations above a few GeV~\cite{aharonianNew}. We take the following approximation for the proton-proton inelastic cross section~\cite{aharonianNew} in millibarnes:}
\begin{equation}
\sigma_{\rm pp}^{\rm inel}(E_{\rm p})\approx 
       (34.3 + 1.88 L +0.25 L^2) \left(1-\left(\frac{E_{\rm th}}{E_{\rm p}} \right)^4\right)^2,
  \label{eq:pion_cross}
\end{equation}

\begin{figure}[t]
	\begin{centering}
		\includegraphics[width=\columnwidth]{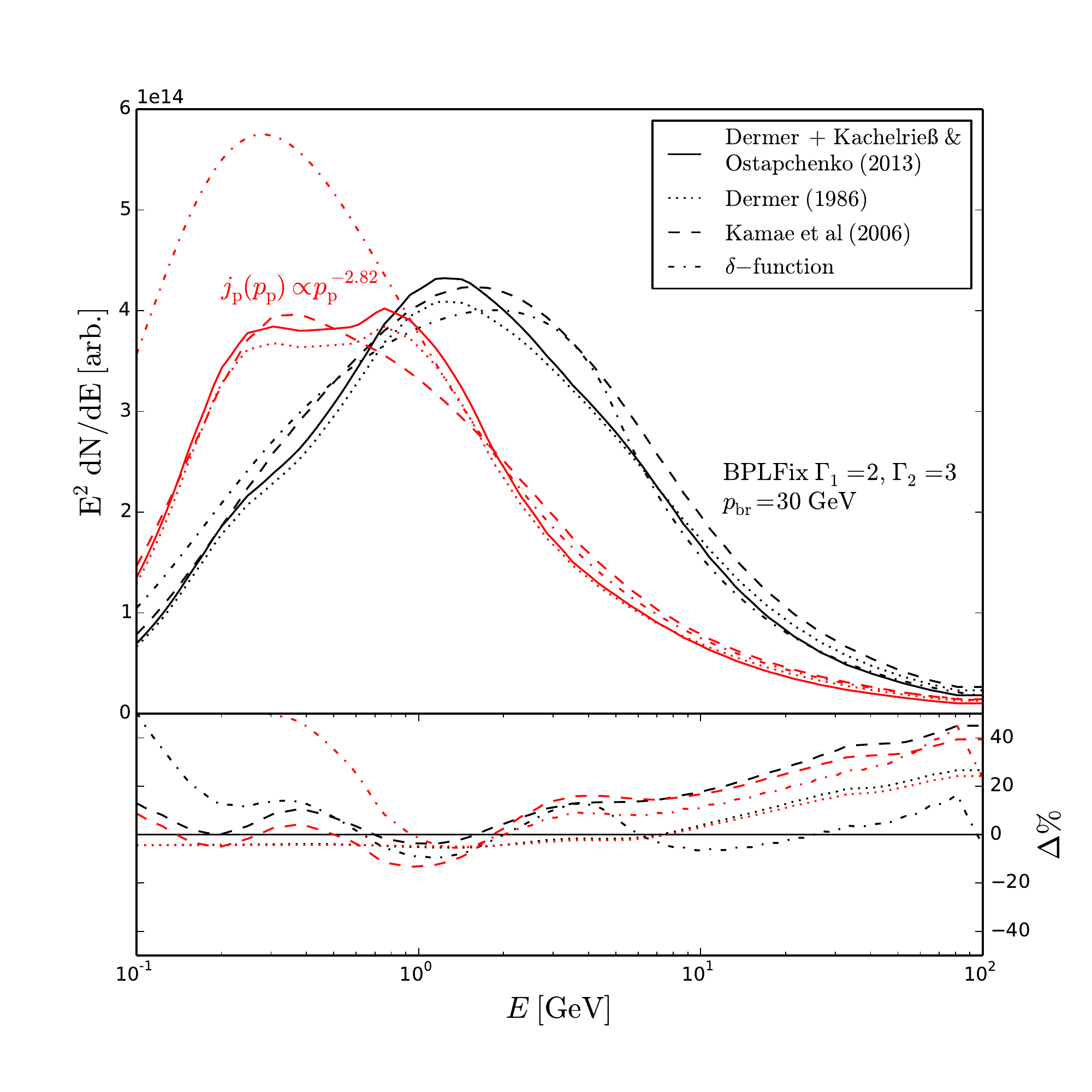}
	\end{centering}
\caption{Model variations in the $\gamma$-ray spectral energy distribution for cosmic-ray proton spectra following (in black) a broken power-law spectrum with $\Gamma_1=2, \Gamma_2=3$, and $p_{\rm br}=30$~GeV (see Eq.~(\ref{eqn:BPL})) and, in red, a flat power-law of index 2.82 representative of the `sea' of Galactic protons. }
\label{fig:pion_decay}
\end{figure}

\rev{where $L=\ln(E_{\rm p}/1~\rm{TeV})$ and $E_{\rm th}=m_{\rm p}+2m_{\pi}+m_\pi^2/(2m_{\rm p})$ is the pion production threshold, below which the inelastic cross section is zero. This provides a reasonable estimate for many cases, but as can be seen in Figure~\ref{fig:pion_decay}, it does not provide an adequate representation of the near-threshold behavior ($T_{\rm p} \lesssim 1$~GeV). Besides integrating over the full range of proton energies (as opposed to approximating with a $\delta$-function) the core difference between this simplified approach and the more sophisticated calculations is a detailed parametrization of the inclusive production cross section and pion multiplicities at low energies, and sometimes Monte Carlo interpolation at high energies.}

Below a few GeV, light hadronic states decaying through $\pi^0$'s provide the main contribution, primarily from the $\Delta (1232)$ resonance.  As the proton energy increases, heavier resonances become more important as well as secondary photons from $\eta$ decays.  The Dermer model includes the $\Delta(1232)$ using Stecker's isobar model~\cite{stecker1971cosmic} at low-energies with linear interpolation between 3 and 7 GeV to the scaling model of Stephens and Badhwar~\cite{stephens}.  \rev{We note that this model relies on cross-sections originally compiled by Stecker in Ref.~\cite{Stecker1973a}.} At higher energies, however, this model violates the Feynman scaling hypothesis, where $E \rm{d}\sigma/\rm{d}^3p$ becomes independent of the center of mass energy $s$ for $s\gg m_{\rm p}^2$.  Kamae et al~\cite{Kamae2006} instead relies on parameterizations of Monte Carlo simulations in addition to corrections for the $\Delta(1232)$, the $N(1600)$ cluster of resonances, diffractive processes, non-scaling effects, and scaling violations which provides a better fit to high-energy observations than Dermer.  The mixed DKO~\cite{Kachelriess:2012} model used in this paper combines simulation/parametrization approaches by interpolating to results from event generator QGSJET-II at energies above $30$~GeV providing a better fit to available high-energy collider data. When fitting a proton spectrum to $\gamma$-ray data, Dermer provides the best fit below 1 GeV, but underestimates the higher-energy spectrum.  Kamae et al has the opposite behavior, matching above 1 GeV, but overproducing photons below. The mixed model provides good fits in both regimes, and hence is the model of choice here.

In the top panel of Figure~\ref{fig:pion_decay} we show in black the $\gamma$-ray spectrum resulting from the fixed broken power-law (BPLFix) model of Eq.~\ref{eqn:BPL} with $\Gamma_1=2, \Gamma_2=3$, and $E_{\rm br}=30$~GeV as well as the background Galactic protons in red following a flat power law with index $\Gamma=2.82$ for each of the four models.  Note that the relative normalization for each of the {\tt Galprop} models is correct while the $\delta$-function case renormalized to match DKO at 2~GeV.  In the lower panel we show the fractional variation in the spectral energy distributions of each model with respect to DKO.  The two most important factors for an analysis of the Galactic center excess are the position and width of the spectral peak.  The models which include the detailed low-energy characterization of Dermer produce the sharpest peak while that of Kamae et al is slightly broadened and peaks at 50\% higher energy for the BPLFix model.  This implies that the $\pi^0$ spectrum using Kamae et al requires slightly softer low and high-energy spectral indices than those of Dermer in order to match the GCE with a broken power law proton spectrum.  Of general interest, but less importance to our analysis is the significant variation in the predicted Galactic background spectrum, where two distinctive peaks are seen in Dermer models compared to only one in the other two.  \rev{It is clear that the $\delta$-function approximation does not accurately characterize the spectrum below $\approx 1$~GeV, and is hence not suitable for calculating spectra over GCE energies.}

\end{appendices}

\bibliography{galcenter}
\end{document}